\documentclass[12pt]{article}
\usepackage{graphicx,amsfonts,amsmath,amssymb,mathrsfs,url}

\title{Paradoxical Reflection in Quantum Mechanics}

\author{ 
Pedro L.\ Garrido\footnote{Departamento de Electromagnetismo y
    F\'\i sica de la Materia, 
    Institute Carlos I for Theoretical and Computational Physics,
    Facultad de Ciencias, Universidad de Granada, 18071 Granada, Spain. E-mail:
    garrido@onsager.ugr.es},
Sheldon Goldstein\footnote{Departments of Mathematics, Physics and
     Philosophy,  Rutgers University,  
     110 Frelinghuysen Road, Piscataway, NJ 08854-8019, USA.
     E-mail: oldstein@math.rutgers.edu},\\
Jani Lukkarinen\footnote{Department of Mathematics and Statistics,
     University of Helsinki, P.O. Box 68,
     FI-00014 Helsingin yliopisto, Finland.
     E-mail: jani.lukkarinen@helsinki.fi}, and
Roderich Tumulka\footnote{Department of Mathematics,
     Rutgers University,
     110 Frelinghuysen Road, Piscataway, NJ 08854-8019, USA.
     E-mail: tumulka@math.rutgers.edu}
}
\date{May 2, 2011}

\newcommand{\be}{\begin{equation}}
\newcommand{\ee}{\end{equation}}
\renewcommand{\Re}{\mathrm{Re}}
\renewcommand{\Im}{\mathrm{Im}}
\newcommand{\norm}[1]{\left\Vert #1 \right\Vert}
\newcommand{\CCC}{\mathbb{C}}
\newcommand{\ZZZ}{\mathbb{Z}}
\newcommand{\RRR}{\mathbb{R}}
\newcommand{\rme}{\mathrm{e}}
\newcommand{\rmd}{\mathrm{d}}
\newcommand{\ci}{\mathrm{i}}
\newcommand{\width}{\sigma}
\newcommand{\vep}{\varepsilon}

\newcommand{\refl}{\mathrm{refl}}
\newcommand{\tra}{\mathrm{tra}}
\newcommand{\inc}{\mathrm{in}}
\newcommand{\cl}{\mathrm{cl}}
\newcommand{\qu}{\mathrm{qu}}

\newtheorem{Thm}{Theorem}
\newtheorem{Corol}{Corollary}

\newcommand{\z}[1]{{#1}}
\newcommand{\y}[1]{{#1}}
\newcommand{\x}[1]{{#1}}
\newcommand{\xj}[1]{#1}

\begin{document}
\maketitle

\begin{abstract}
This article concerns a phenomenon of elementary quantum mechanics that
is quite counter-intuitive, very non-classical, and apparently not 
widely known: a quantum particle can get reflected at a downward potential
step. In contrast, classical particles get reflected only at upward
steps. The conditions for this effect are that the wave length is much 
greater than the width of the potential step and the kinetic energy of
the particle is much smaller than the depth of the potential step.
This phenomenon is suggested by non-normalizable solutions to the time-independent Schr\"odinger equation, and we present evidence, numerical and mathematical, that it is also indeed predicted by the time-dependent Schr\"odinger equation. Furthermore, this paradoxical reflection effect suggests, and we confirm mathematically, that
a quantum particle can be trapped for a long time
(though not forever) in a region surrounded by downward potential steps, that is, on a plateau. 

\medskip

\end{abstract}

\section{Introduction}

\begin{figure}[ht]\begin{center}\includegraphics[width=.4 \textwidth]{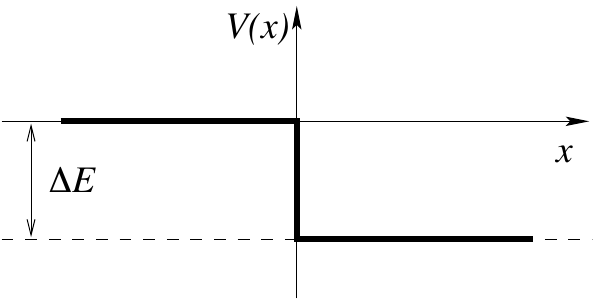}\end{center}\caption{A potential $V(x)$ containing a downward step}\label{fig:step}\end{figure}

Suppose a quantum particle moves towards a sudden drop of potential as in
Figure \ref{fig:step}, with the particle arriving from the left. 
Will it accelerate or be reflected? A classical particle is certain
to accelerate, but a quantum particle has a chance to be reflected. That
sounds paradoxical because the particle turns around and returns to the
left under a force pointing to the right! Under suitable conditions, reflection even becomes close to certain. This non-classical, counter-intuitive quantum phenomenon we call ``paradoxical
reflection,'' or, when a region is surrounded by downward potential steps, ``paradoxical confinement''---where ``paradoxical'' is understood in the sense of
``counter-intuitive,'' not ``illogical.'' It can be derived easily using the
following simple reasoning. 

Suppose the particle moves in 1 dimension, and the potential is a rectangular step as in Figure \ref{fig:step},
\begin{equation}\label{V}
  V(x) = -\Delta E \: \Theta(x)
\end{equation}
with $\Theta$ the Heaviside function and $\Delta E \geq 0$. A wave packet
coming from the left gets partially reflected at the step and partially
transmitted. The size of the reflected and the transmitted packets can be
determined by a standard textbook method of calculation (e.g.,
\cite{CTDL,LL}), the \emph{stationary analysis}, replacing the wave \z{packet
by a plane wave of energy $E$} and solving the stationary Schr\"odinger equation. The
transmitted and reflected probability currents, divided by the incoming
current, yield the \emph{reflection} and \emph{transmission coefficients}
$R \geq 0$ and $T\geq 0$ with $R+T=1$. We give the results in Section
\ref{sec:rect} and observe two things: First, $R\neq 0$, implying that
partial reflection occurs although the potential step is
\emph{downward}. Second, $R$ even converges to 1, so that reflection
becomes nearly certain, as the ratio $E/\Delta E$ goes to zero. Thus,
paradoxical reflection can be made arbitrarily strong by a suitable
choice of parameters
(e.g., for sufficiently big $\Delta E$ if $E$ is kept fixed). 

If it sounds incredible that a particle can be repelled by a downward 
potential step, the following fact may add to the amazement. As derived in
\cite[p.~76]{LL}, the reflection coefficient does not depend on whether the
incoming wave comes from the left or from the right (provided the total
energy and the potential are not changed). Thus, a downward step
yields the same reflection coefficient as an upward step. (But keep
in mind the difference between an upward step and a downward step 
that at an upward step, also energies below the height of the step are
possible for the incoming particle, a case in which reflection is certain, $R=1$.)

To provide some perspective, it may be worthwhile to point to
some parallels with quantum tunneling: there, the probability of a quantum
particle passing through a potential barrier is positive even in cases in
which this is impossible for a classical particle. In fact, paradoxical
reflection is somewhat similar to what could be called
\emph{anti-tunneling}, the effect that a quantum particle can have positive
probability of being reflected by a barrier so small that a classical
particle would be certain to cross it. Paradoxical reflection is less surprising
when we think of a wave being reflected from a potential step, and
more surprising from the particle point of view. It is sometimes, though
apparently not frequently, pointed out in textbooks 
\cite[p.~84]{G}, \cite[p.~197-8]{ER}.

The goal of this article is to address the following questions: Is
paradoxical reflection a real physical phenomenon or an artifact of
mathematical over-simplification? (We will look at numerical and
rigorous mathematical results.) How does it depend on the
parameters of the situation: the width $L$ (see Figure
\ref{fig:softstep}) and the depth $\Delta E$ of the
potential step, the wave length $\lambda$ and the width $\sigma$ of the
incoming wave packet? Why does this phenomenon not occur in the classical
regime? That is, how can it be that classical mechanics is a limit of
quantum mechanics if paradoxical reflection occurs in the latter but not
the former? And, could one use this phenomenon in principle for
constructing a particle trap? In spring 2005, these questions gave rise to
lively and contentious discussions between a number of physics
researchers visiting the Institut des Hautes \'Etudes Scientifiques near
Paris, France; these discussions inspired the present article.

\section{Stationary Analysis of the Rectangular Step}\label{sec:rect}

We begin by providing more detail about the stationary analysis of the
rectangular step \eqref{V}, considering the time-independent Schr\"odinger
equation ($m = \text{mass}$) 
\begin{equation}\label{TISE}
  E \psi(x) = -\tfrac{\hbar^2}{2m} \psi''(x) + V(x) \, \psi(x)\,.
\end{equation}
This can be solved in a standard way: for $x<0$, let $\psi$ be a
superposition of an incoming wave $\rme^{\ci k_1x}$ and a reflected wave $B
\rme^{-\ci k_1x}$, while for $x>0$, let $\psi$ be a transmitted wave
$A\rme^{\ci k_2x}$, with a possibly different wave number $k_2$. Indeed,
from \eqref{TISE} we obtain that  
\begin{equation}\label{ks}
  k_1 = \sqrt{2mE}/\hbar \,, \quad k_2 = \sqrt{2m(E+\Delta E)}/\hbar\,.
\end{equation}
The value $E\geq 0$ is the kinetic energy associated with the incoming wave.
The coefficients $A$ and $B$ are determined by continuity of $\psi$ and its derivative $\psi'$ at $x=0$ to be
\begin{equation}\label{AB}
  A = \frac{2k_1}{k_1 + k_2}\,, \quad B = \frac{k_1 - k_2}{k_1 + k_2}\,.
\end{equation}
The \emph{reflection} and \emph{transmission coefficients} $R$ and $T$ are
defined as the quotient of the quantum probability current $j=(\hbar/m)
\Im(\psi^*\psi')$ associated with the reflected respectively transmitted
wave divided by the current associated with the incoming wave, 
\begin{equation}\label{RTdef}
  R = \frac{|j_\refl|}{j_\inc} \,, \quad T = \frac{j_\tra}{j_\inc}\,.
\end{equation}
\z{Noting that $j_\tra = \hbar k_2 |A|^2/m$, $j_\refl = -\hbar k_1
|B|^2/m$, $j_\inc = \hbar k_1/m$, we find that}
\begin{equation}\label{RT}
  R = |B|^2 = 1- \frac{k_2}{k_1} |A|^2\,,\quad T = \frac{k_2}{k_1} |A|^2\,.
\end{equation}
Note that both $R$ and $T$ lie in the interval $[0,1]$, and that $R+T=1$.
By inserting \eqref{AB} into \eqref{RT}, we obtain
\begin{equation}\label{Rr}
  R = \frac{(k_1+k_2)^2 -4k_1k_2}{(k_1 + k_2)^2} = 
  \frac{(k_2-k_1)^2}{(k_1 + k_2)^2}
\end{equation}
and make two observations: First, $R\neq 0$\z{, implying that reflection
occurs,}  if $k_1 \neq k_2$, which is the
case as soon as $\Delta E \neq 0$. Second,
$R$ even converges to 1, so that reflection becomes nearly certain, as the
ratio $r:= E/\Delta E$ tends to zero; that is because 
\begin{equation}
  R = \Biggl( \frac{k_2-k_1}{k_2+k_1} \Biggr)^2 = \Biggl( \frac{\sqrt{E+\Delta E}
  -\sqrt{E}}{\sqrt{E+\Delta E}+\sqrt{E}} \Biggr)^2 = \Biggl( \frac{\sqrt{r+1} - \sqrt{r}}
  {\sqrt{r+1}+\sqrt{r}} \Biggr)^2 \to 1\,,
\end{equation}
since both the numerator and the denominator tend to 1 as $r \to 0$. This
is the simplest derivation of paradoxical reflection. 

The effect possesses an analog in wave optics. The refractive index, which may vary with the position $x$, plays a role similar to the potential (e.g., in that it influences the speed of wave propagation), and changes suddenly at a surface between different media, say, between water and air. Light can be reflected at the surface on both sides; in particular, light coming from the water (the high-index region) can be reflected back into the water.

\section{Soft Step}
\label{sec:softstep}

\begin{figure}[ht]\begin{center}\includegraphics[width=.4 \textwidth]{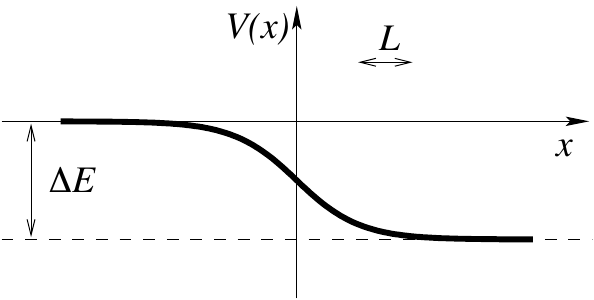}\end{center}\caption{A potential containing a soft step}\label{fig:softstep}\end{figure}

For a deeper analysis of the effect, we will gradually consider increasingly more
realistic models. In this section, we consider a soft (or smooth, i.e.,
differentiable) potential step, as in Figure \ref{fig:softstep}, for which
the drop in the potential is not infinitely rapid but takes place over some
distance $L$. The result will be that paradoxical reflection exists also
for soft steps, so that the effect is not just a curious feature of
rectangular steps (which could not be expected to ever occur in
nature). Another result concerns how the effect depends on the width $L$ of
the step. 

To study this case it is useful to consider the explicit function
\begin{equation}\label{Vs}
  V(x) = -\frac{\Delta E}{2} \Bigl( 1 + \tanh \frac{x}{L} \Bigr)\,,
\end{equation}
depicted in Figure \ref{fig:softstep}. (Recall that $\tanh = \sinh /\cosh$
converges to $\pm 1$ as $x \to \pm \infty$.) The reflection coefficient for
this potential can be calculated again by a stationary analysis, obtaining
from the time-independent Schr\"odinger equation \eqref{TISE} solutions
$\psi(x)$ which are asymptotic to $\rme^{\ci k_1x} + B \rme^{-\ci k_1x}$ as
$x \to -\infty$ and asymptotic to $A \rme^{\ci k_2x}$ as $x \to \infty$,
i.e., $\lim_{x \to \infty} (\psi(x) - A \rme^{\ci k_2x}) = 0$. The
calculation is done in \cite[p.~78]{LL}: The values $k_1$ and $k_2$ are
again given by \eqref{ks}, and the reflection coefficient turns out to be 
\begin{equation}\label{Rs}
  R = \Biggl( \frac{\sinh \bigl(\frac{\pi}{2}(k_2-k_1)L \bigr)}
  {\sinh \bigl( \frac{\pi}{2}(k_2+k_1)L\bigr)} \Biggr)^2\,.
\end{equation}
From this and \eqref{ks} we can read off that, again, $R \neq 0$ for
$\Delta E \neq 0$, and $R \to 1$ as $E \to 0$ while $\Delta E$ and $L$ are
fixed (since then $k_1\to 0$, $k_2\to \sqrt{2m\Delta E}/\hbar$, so both the numerator and the denominator tend to $\sinh 
(\frac{\pi}{2} \sqrt{2m\Delta E} L/\hbar)$). 
As $\Delta E \to \infty$ while $E$ and $L$ are fixed, $R \to
\exp(-2\pi\sqrt{2mE}L/\hbar)$ because for large arguments $\sinh \approx
\tfrac{1}{2} \exp$. 

In addition, we can keep $E$ and $\Delta E$ fixed and see how $R$ varies
with $L$: In the limit $L \to 0$, \eqref{Rs} converges to \eqref{Rr}
because $\sinh(\alpha L) \approx \alpha L$ for $L \ll 1$ and fixed
$\alpha$; this is what one would expect when the step becomes sharper and
\eqref{Vs} converges to \eqref{V}. In the limit $L \to \infty$, $R$
converges to $0$ because for fixed $\beta >\alpha >0$ 
\begin{equation}
  \frac{\sinh (\alpha L)}{\sinh (\beta L)} = 
  \frac{\rme^{\alpha L} - \rme^{-\alpha L}}{\rme^{\beta L} - \rme^{-\beta L}} =
  \frac{\rme^{(\alpha-\beta)L}-\rme^{(-\alpha-\beta)L}} {1-\rme^{-2\beta L}} \to 0\,,
\end{equation}
as the numerator tends to 0 and the denominator to 1. Thus, paradoxical
reflection disappears for large $L$; in other words, it is crucial for the
effect that the drop in the potential is \emph{sudden}. 

Moreover, \eqref{Rs} is a decreasing function of $L$, which means that
reflection will be the more probable the more sudden the drop in the
potential is. To see this, let us check that for $\beta>\alpha>0$ and $L>0$
the function $f(L) = \sinh(\alpha L)/\sinh(\beta L)$ is decreasing: 
\begin{equation}
  \frac{\rmd f}{\rmd L} = \frac{\alpha \cosh(\alpha L) \sinh(\beta L) - 
  \beta \sinh(\alpha L) \cosh(\beta L)}{\sinh^2(\beta L)} < 0
\end{equation}
because
\begin{equation}
  \frac{\alpha}{\tanh \alpha} < \frac{\beta}{\tanh \beta}\,,
\end{equation}
as $x/\tanh x$ is increasing for $x>0$.

How about soft steps with other shapes than that of the $\tanh$ function?
Suppose that the potential $V(x)$ is a continuous, monotonically decreasing
function such that $V(x) \to 0$ as $x \to -\infty$ and $V(x) \to -\Delta E$
as $x \to +\infty$. To begin with, we note that the fact, mentioned in the
introduction, that the reflection coefficient is the same for particles
coming from the left or from the right, still holds true for such a general
potential \cite[p.~76]{LL}. This suggests that paradoxical
reflection occurs also for general potential steps. Unfortunately, we do
not know of any general result on lower bounds for the reflection
coefficient $R$ that could be used to establish  paradoxical
reflection in this generality. However, an upper bound is known \cite[eq.~(82)]{Visser},
according to which $R$ is less than or equal to the reflection coefficient
\eqref{Rr} of the rectangular step. This agrees with our observation in the
previous paragraph that reflection is the more likely the sharper the
step.

\section{Wave Packets}

Another respect in which we can be more realistic is by admitting that the
wave function with which a quantum particle reaches a potential step is not
an infinitely-extended plane wave $\rme^{\ci k_1x}$ but in fact a wave
packet of finite width $\sigma$, for example a Gaussian wave packet 
\begin{equation}\label{Gaussian}
  \psi_\inc(x) = G_{\mu,\sigma}(x)^{1/2} \, \rme^{\ci k_1x}
\end{equation}
with $G_{\mu,\sigma}$ the Gauss function with mean $\mu$ and variance
$\sigma^2$,
\begin{equation}
G_{\mu,\sigma}(x) = \frac{1}{\sqrt{2\pi}\sigma} \rme^{-(x-\mu)^2/2\sigma^2}\,.
\end{equation}
Suppose this packet arrives from the left and evolves in the
potential $V(x)$ according to the time-dependent Schr\"odinger equation 
\begin{equation}\label{TDSE}
  \ci\hbar\frac{\partial \psi}{\partial t}(x,t) =  
  -\frac{\hbar^2}{2m} \frac{\partial^2 \psi}{\partial x^2} (x,t) + V(x) \, \psi(x,t)\,.
\end{equation}
Ultimately, as $t \to \infty$, there will be a reflected packet
$\psi_\refl$ in the region $x<0$ moving to the left and a transmitted
packet $\psi_\tra$ in the region $x>0$ moving to the right, and thus the
reflection and transmission probabilities are 
\begin{equation}\label{RTprob}
  R= \|\psi_\refl\|^2 \,, \quad T = \|\psi_\tra\|^2\,,
\end{equation}
with $\|\psi\|^2=\int_{-\infty}^\infty |\psi(x)|^2\, \rmd x$.

Because of the paradoxical feel of paradoxical reflection, one might
suspect at first that the effect does not exist for wave packets but is merely an artifact
of the stationary analysis. We thus address, in this section, the question
as to how wave packets behave, and whether the reflection probability
\eqref{RTprob} agrees with the reflection coefficient discussed earlier. We
begin with the numerical evidence confirming paradoxical reflection.

\subsection{Numerical Simulation}

\begin{figure}[ht]
\begin{center}
\includegraphics[width=12cm]{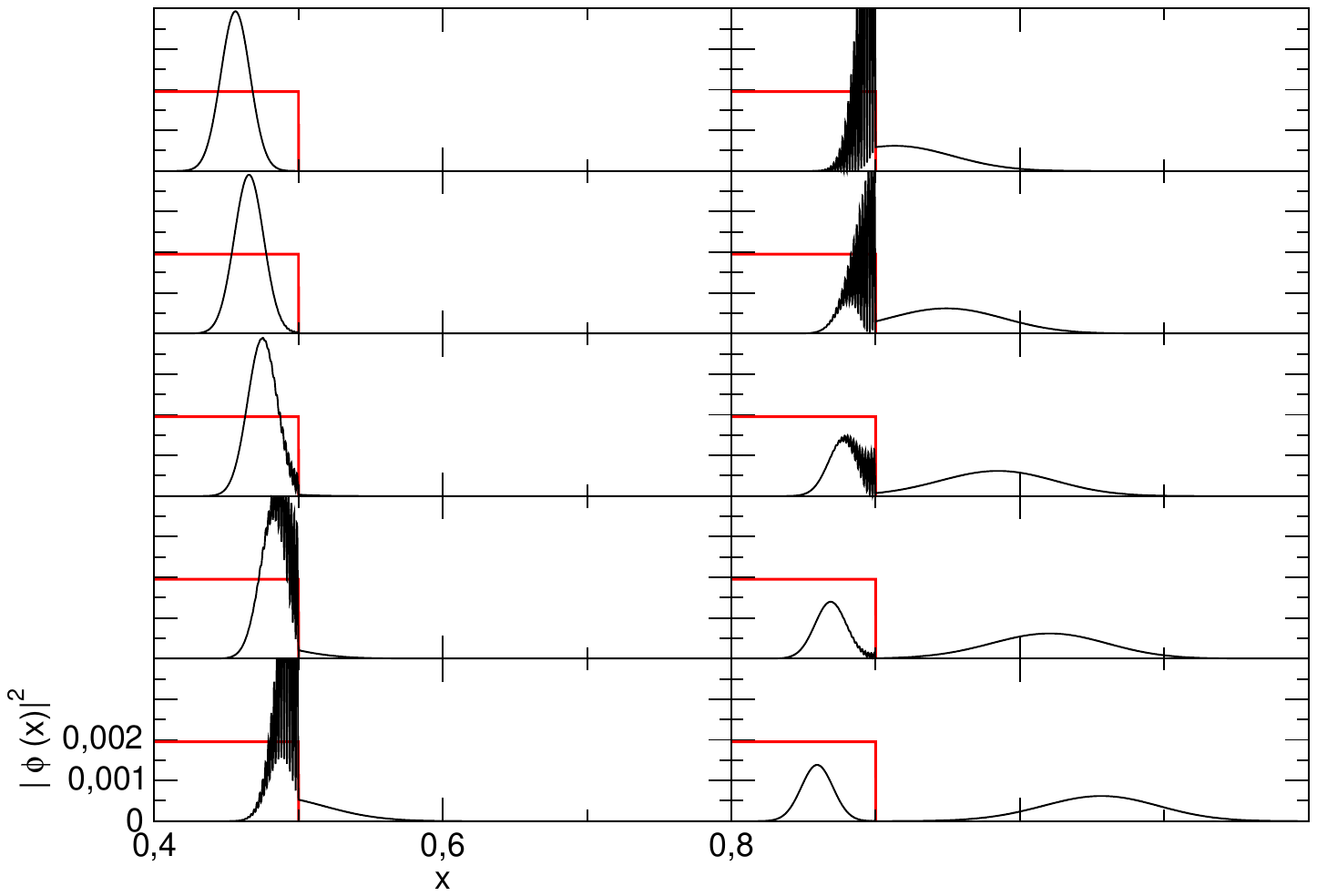}
\end{center}
\caption{Numerical simulation of the time-dependent Schr\"odinger equation for the \y{hard step potential \eqref{V}.} The picture shows ten snapshots of $|\psi|^2$ (black lines) at different times before, during, and after passing the potential step. (Order: left column top to bottom, then right column top to bottom.) \y{The additional lines in the figures depict the potential in
arbitrary units.} It can clearly be seen that there is a transmitted wave packet and, ``paradoxically,'' a reflected wave packet.
The initial wave function is a Gaussian wave packet centered \y{at $x=0.4$ with $\sigma=0.01$ and $k_0=500\pi$. The simulation assumes
infinite potential walls at $x=0$ and $x=1$. 
The step height is $\Delta E = 15 \, E$, 
and the $x$-interval is resolved with a linear mesh of $N=10^4$ points. The snapshots are taken at times
$6,7,8,\ldots,15$ in appropriate time units.}}
\label{fig:numerical}
\end{figure}

A numerical simulation of a wave packet partly reflected from a \y{(hard)} downward step is shown in Figure \ref{fig:numerical}. The simulation starts with a Gaussian wave packet moving to the right and initially located on the left of the potential step. After passing the step, there remain two wave packets, no longer of exactly Gaussian shape, one continuing to move to the right and the other, reflected one returning to the left. For the choice of parameters in this simulation, the transmitted and reflected packet are of comparable size, thus providing evidence that there can be a substantial probability of reflection at a downward step (even for wave packets of finite width). That is, the numerical simulation confirms the prediction of the stationary analysis.

\subsection{But Is It for Real?}\label{sec:rigor}

We now point out how rigorous mathematics confirms paradoxical reflection
as a consequence of the Schr\"odinger equation. We thus exclude the
possibility that it was merely numerical error that led to the appearance of paradoxical
reflection for wave packets.

\begin{figure}[t]
\begin{center}
\includegraphics[width=12cm]{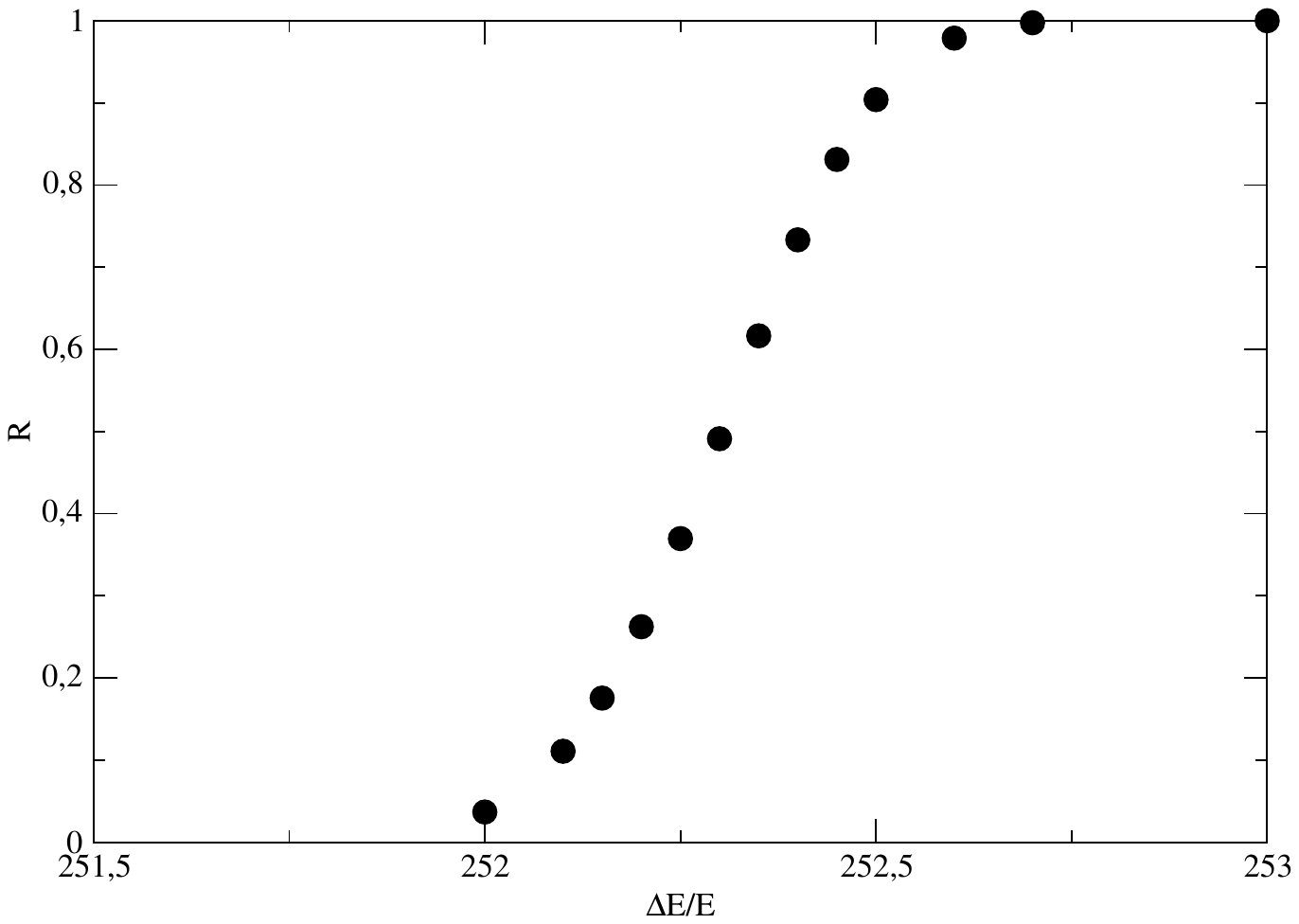}
\end{center}
\caption{An example of how numerical error may lead to wrong predictions. \y{The simulation shown in Figure~\ref{fig:numerical} was repeated with a soft step potential \eqref{Vs} with $L=0.005$ for different values of the step height $\Delta E$. The plot shows the numerical values for the reflection probability $R= \|\psi_\refl\|^2$. These values cannot be correct; indeed, for the parameters used in this simulation (see below), $R$ cannot get close to 1 but must stay between 0 and $10^{-17}$ for every $\Delta E>0$. The simulation used a standard algorithm for simulating the Schr\"odinger equation \cite{GSS}, a grid of $N=10^4$ sites, and as the initial wave function a Gaussian packet with parameters $k_1=400\pi$, $x_0=0.4$, $\sigma=0.005$. The bound of $10^{-17}$ follows from Eq.~\eqref{RRk} below and the fact that the reflection coefficient \eqref{Rs} is bounded by $\exp(-2\pi\sqrt{2mE}L/\hbar)=\exp(-2\pi k_1 L)$, which here is $\exp(-4\pi^2)<10^{-17}$.}}
\label{fig:error}
\end{figure}

Do not think the worry that numerical errors may lead to the wrong behavior of a 
wave packet was paranoid: There are cases in which exactly this happens.
For example, when we carried out a simulation of the evolution of a wave packet in a soft step potential (i.e., the same situation as in Figure~\ref{fig:numerical} but with the hard step \eqref{V} replaced by the soft step \eqref{Vs}) we obtained completely wrong outcomes for the reflection probabilities; see Figure~\ref{fig:error}.


We return to the mathematics of paradoxical reflection. 
The rigorous mathematical analysis of scattering problems of this type
is a fairly complex and subtle topic.  The main techniques and results 
(also for higher dimensional problems) are described in
\cite{RS79,RS78}, and  the mathematical results relevant to potentials of 
the step type can be found in \cite{SD78}. 
The reflection probability $R$ of eq.~\eqref{RTprob} is given in terms of the plane wave reflection coefficients $R(k_1)$ by the following formula, expressing exactly what one would intuitively expect: 
\begin{equation}\label{RRk}
  R = \int_0^\infty \rmd k_1 \, R(k_1) \, |\widehat{\psi}_\inc(k_1)|^2 \,.
\end{equation}
The same formula holds with all $R$'s replaced by $T$'s. In
\eqref{RRk}, $R(k_1)$ is given by the stationary analysis, as in
\z{\eqref{Rr} or}
\eqref{Rs}, with $k_2$ expressed in terms of $k_1$ and $\Delta E$,
$k_2 =\sqrt{k_1^2 + 2m\Delta E/\hbar^2}$; and $\widehat{\psi}_\inc(k_1)$ is the
Fourier transform of the incoming wave packet $\psi_\inc(x)$. 

(In brackets: To be precise, the incoming packet $\psi_{\inc}(x,t)$ is
defined as the free asymptote of $\psi(x,t)$ for $t \to -\infty$, i.e.,
$\psi_{\inc}(x,t)$ evolves without the potential, 
\begin{equation}\label{fTDSE}
  \ci\hbar\frac{\partial \psi_\inc}{\partial t}(x,t) =  
  -\frac{\hbar^2}{2m} \frac{\partial^2 \psi_{\inc}}{\partial x^2} (x,t)\,,
\end{equation}
and
\begin{equation}
  \lim_{t\to-\infty} \|\psi_\inc(\cdot,t) - \psi(\cdot,t)\| = 0\,.
\end{equation}
Similarly, $\psi_\refl + \psi_\tra$ is the free asymptote of $\psi$ for $t
\to +\infty$. When we write $\psi_\inc(x)$, we mean to set $t=0$; note,
however, that \eqref{RRk} actually does not depend on the $t$-value as, by
the free Schr\"odinger equation \eqref{fTDSE}, $\widehat{\psi}_\inc(k,t) =
\exp(-it\hbar k^2/2m)\, \widehat{\psi}_\inc(k,0)$ and thus
$|\widehat{\psi}_\inc(k,t)|^2 = |\widehat{\psi}_\inc(k,0)|^2$. Since we
assumed that the incoming wave packet comes from the left, $\psi_\inc$ is a
``right-moving'' wave packet consisting only of Fourier components with
$k\geq 0$.) 

From \eqref{RRk} we can read off the following: If the incoming wave packet
consists only of Fourier components $k_1$ for which $R(k_1) >
1-\varepsilon$ for some (small) $\varepsilon >0$, then also $R> 1-
\varepsilon$. More generally, if the incoming wave packet consists
\emph{mainly} of Fourier components with $R(k_1) >1-\varepsilon$, that is,
if the proportion of Fourier components with $R(k_1) >1-\varepsilon$ is 
\begin{equation}
  \int_0^\infty \rmd k_1 \, \Theta \bigl( R(k_1)-(1-\varepsilon) \bigr) 
  \,  |\widehat{\psi}_\inc(k_1)|^2 = 1-\delta\,,
\end{equation}
then $R>1-\varepsilon-\delta$ because 
\[
 \int_0^\infty \rmd k_1 \, R(k_1) \,  |\widehat{\psi}_\inc(k_1)|^2 \geq 
 \int_0^\infty \rmd k_1 \, R(k_1)\, \Theta \bigl( R(k_1)-(1-\varepsilon) \bigr) 
 \,  |\widehat{\psi}_\inc(k_1)|^2 \geq
\]
\[
 \geq \int_0^\infty \rmd k_1 \, (1-\varepsilon) \, \Theta \bigl( R(k_1)-(1-\varepsilon) \bigr) 
 \,  |\widehat{\psi}_\inc(k_1)|^2 = (1-\varepsilon)(1-\delta) 
 > 1-\varepsilon-\delta \,.
\] 

Therefore, whenever the stationary analysis predicts paradoxical
reflection for certain parameters and values of $k_1$, then also wave
packets consisting of such Fourier components will be subject to
paradoxical reflection.

\section{Parameter Dependence}

Let us summarize and be explicit about how the reflection probability $R$
from a downward potential step depends on the parameters of the situation:
the mean wave number $k_1$ and  the width $\sigma$ of the incoming wave
packet, and the depth $\Delta E$ and width $L$ of the potential step. We
claim that $R$ is close to 1 in the parameter region with 
\begin{subequations}\label{para}
\begin{align}
  \frac{1}{k_1} &\gg L \label{parakL}\\
  \Delta E & \gg \frac{\hbar^2 k_1^2}{2m}= E \label{paraDeltaE}\\
  \sigma & \gg \frac{1}{k_1} \label{parasigma}\,.
\end{align}
\end{subequations}
Note that $1/k_1$ is (up to the factor $2\pi$) the (mean) wave length $\lambda$.

\begin{figure}[ht]
\begin{center}
\includegraphics[angle=-90,width=.4 \textwidth]{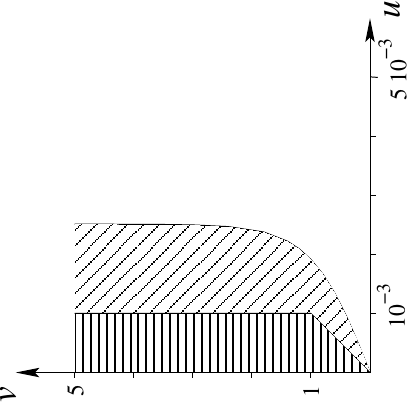}
\end{center}
\caption{The region (shaded) in the plane of the parameters $u$ and $v$, defined in \eqref{uvdef}, in which the reflection probability \eqref{Ruv} exceeds 99 percent. The horizontally shaded subset is the region in which condition \eqref{uv1000} holds.}
\label{fig:parameter}
\end{figure}

To derive this claim from \eqref{Rs} and \eqref{RRk}, consider first the
case $\sigma \to \infty$ of a very very wide packet. For such a packet, its
Fourier transform is very sharply peaked at $k_1$. The reflection
coefficient $R$ given by \eqref{Rs} depends on the parameters
$k_1, L, \Delta E, m$ only in the dimensionless combinations 
\begin{equation}\label{uvdef}
  u= \tfrac{\pi}{2} k_1 L\,, \quad v= \tfrac{\pi}{2} \sqrt{2m\Delta E} L/\hbar\,,
\end{equation}
that is, 
\begin{equation}\label{Ruv}
  R=R(u,v) = \Biggl( \frac{\sinh (\sqrt{u^2+v^2}-u)}
  {\sinh (\sqrt{u^2+v^2}+u)} \Biggr)^2 \,. 
\end{equation}
Figure \ref{fig:parameter} shows the region in the $uv$ plane in which $R>0.99$. 
As one can read off from the figure, for $(u,v)$ to lie in that region, it
is sufficient, for example, that
\begin{equation}\label{uv1000}
u<10^{-3}\text{ and }v>10^3 u\,. 
\end{equation}

More generally, for $R(u,v)$ to be very close to 1 it is sufficient that $u \ll
1$ and $v \gg u$, which means \eqref{parakL} and \eqref{paraDeltaE}.  
To see this, note that
\begin{equation}
\begin{split}
  &\sinh(\sqrt{u^2+v^2}-u) = \sinh (\sqrt{u^2+v^2} + u - 2u) =\\
  & \quad = \:\sinh(\sqrt{u^2+v^2}+u) \cosh(2u) - \cosh(\sqrt{u^2+v^2}+u) \sinh(2u)
\end{split}
\end{equation}
so that
\begin{equation}
  \sqrt{R(u,v)} = \frac{\sinh (\sqrt{u^2+v^2}-u)}
  {\sinh (\sqrt{u^2+v^2}+u)} = \cosh(2u) - \frac{\sinh(2u)}{\tanh (\sqrt{u^2+v^2}+u)}\,.
\end{equation}
\z{Suppose that $u\ll 1$. Then} Taylor expansion to first order in $u$ yields
\begin{equation}\label{srR}
  \sqrt{R(u,v)} \approx 1-\frac{2u}{\tanh v}\,.
\end{equation}
If $v$ is of order 1, this is close to 1 because $u\ll 1$. If, however, $v$
is small, then $\tanh v$ is of order $v$, and \z{the right hand side of
\eqref{srR} is close to 1 when $u/v\ll
1$}. Thus, \z{when \eqref{parakL} and \eqref{paraDeltaE} are satisfied} $\sqrt{R}$ is close to 1, and thus so is $R$. 

Now consider a wave packet that is less sharply peaked in the momentum
representation. If it has width $\sigma$ in position space then, by the
Heisenberg uncertainty relation, it has width of order $1/\sigma$ in
Fourier space. For the reflection probability to be close to one, the wave
packet should consist almost exclusively of Fourier modes that have
reflection coefficient close to one. Thus, every wave number $\tilde{k}_1$
in the interval, say, $[k_1-\frac{10}{\sigma}, k_1+\frac{10}{\sigma}]$
should satisfy \eqref{parakL} and \eqref{paraDeltaE}. This will be the case if
$\frac{10}{\sigma}$ is small compared to $k_1$, or $\sigma\gg 1/k_1$. 
Thus, \eqref{parasigma}, which is merely what is required for
\eqref{Gaussian} to be a good wave packet, i.e., an approximate plane
wave, is a natural condition on $\sigma$ for keeping $R$ close to 1.

\section{The Classical Limit}

If paradoxical reflection exists, then why do we not see it in the classical limit?
On the basis of \eqref{para} we can understand why: Classical mechanics is a good
approximation to quantum mechanics in the regime in which a wave packet
moves in a potential that varies very slowly in space, so that the force
varies appreciably only over distances much larger than the wave length. For paradoxical
reflection, in contrast, it is essential that the length scale of the drop
in the potential be smaller than the wave length. For further discussion of the classical limit of quantum mechanics, see \cite{classical}.

\section{A Plateau as a Trap}

\begin{figure}[ht]\begin{center}\includegraphics[width=.4 \textwidth]{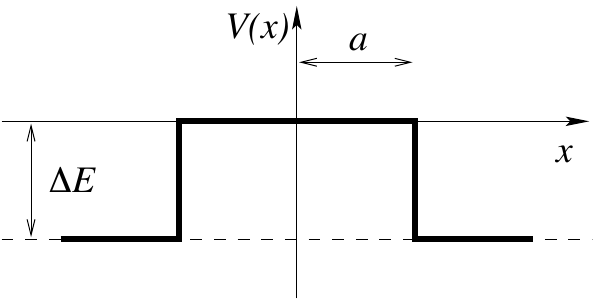}\end{center}\caption{Potential plateau}\label{fig:plateau}\end{figure}

Given that a quantum particle will likely be reflected from a suitable downward potential step, it is obvious that it could be trapped, more or less, in a
region surrounded by such potential steps. 
In other words, also potential plateaus, not only potential valleys, can be
confining.  
To explore this possibility of ``paradoxical confinement,'' we now consider a potential plateau 
\begin{equation}\label{Vp}
  V(x) = -\Delta E \bigl( \Theta(x-a) + \Theta(-x-a) \bigr)\,,
\end{equation}
depicted in Figure \ref{fig:plateau}. 

A particle starting on the plateau could remain there---at least with high
probability---for a very long time, much longer than the maximal time
$\tau_\cl$ that a classical particle with energy $E$ would remain
on the plateau, which is 
\begin{equation}\label{tcl}
  \tau_\cl = a\sqrt{\frac{2m}{E}}\,,
\end{equation}
independently of the height of the plateau. \y{The following theorem, which will be proved in Appendix~\ref{sec:metastable} using the results of Sections \ref{sec:complexenergy}, \ref{sec:packetsonplateau}, and Appendix~\ref{sec:plateigen}, guarantees that paradoxical confinement actually works for sufficiently high plateaus.}

\begin{Thm}\label{thm:confinement1}
Let $a>0$ and choose an initial wave function $\psi_0$ that has $\psi_0(x)=0$ for $|x|>a$ and is normalized but otherwise arbitrary for $|x|\le a$. For every constant $\Delta E>0$, consider the potential $V$, as in \eqref{Vp} and in  Figure~\ref{fig:plateau}, and the time-evolved wave function $\psi_t= \rme^{-\ci H t/\hbar} \psi_0$ (with $H$ denoting the unique self-adjoint extension of $-\tfrac{\hbar^2}{2m}\partial^2/\partial x^2 +V$); we write $\psi_t=\psi_t^{\Delta E}$ to make explicit the dependence on $\Delta E$. During an arbitrarily long time interval $[0,t_0]$ and with arbitrarily small error $\varepsilon>0$, $\psi_t^{\Delta E}$ stays concentrated in the plateau region $[-a,a]$, i.e.,
\be
\int_{-a}^a \bigl| \psi_t^{\Delta E}(x) \bigr|^2 \,\rmd x > 1-\varepsilon 
\quad \text{for all } t \in [0,t_0],
\ee
provided $\Delta E$ is large enough, $\Delta E \geq \Delta E_0(\psi_0,t_0,\varepsilon)$.
\end{Thm}

\y{Given a fixed $\Delta E$, though, the quantum particle does not stay forever in the plateau region. The time it likely remains there} is
of the order $\sqrt{\Delta E/E}\, \tau_\cl$
and is thus much larger than $\tau_\cl$ if the height $\Delta E$ is large enough. In fact, as we shall \y{prove} in the subsequent sections, a quantum particle
starting \y{in the plateau region will leave it, if $\Delta E$ is large enough,} at the rate
$\tau_\qu^{-1}$ with the decay time
\begin{equation}\label{tqu}
  \tau_\qu =  a\frac{\sqrt{2m\Delta E}}{4E} =
  \frac14\sqrt{\frac{\Delta E}{E}}\, \tau_\cl \, .
\end{equation}

The lifetime \eqref{tqu} can be obtained in the following
semi-classical way: Imagine a particle traveling along the plateau with the
speed $\sqrt{2E/m}$ classically corresponding to energy $E$, getting
reflected at the edge with probability $R$ given by \eqref{Rr}, traveling
back with the same speed, getting reflected at the other edge with
probability $R$, and so on. Since the transmission probability $T=1-R$
corresponding to \eqref{Rr} is
\begin{equation}
4\sqrt{E/\Delta E}+ \text{higher powers of }E/\Delta E\,,
\end{equation}
a number of reflections of order
$(\sqrt{E/\Delta E})^{-1}$ should typically be required before
transmission occurs, in qualitative agreement with \eqref{tqu}. In fact,
the transmission probability of $T=4\sqrt{E/\Delta E}$, when small,
corresponds to a decay rate $T/\tau_\cl$ and hence to the decay time
$\tau_\cl/T$ given by \eqref{tqu}.

\begin{figure}[ht]\begin{center}\includegraphics[width=.4 \textwidth]{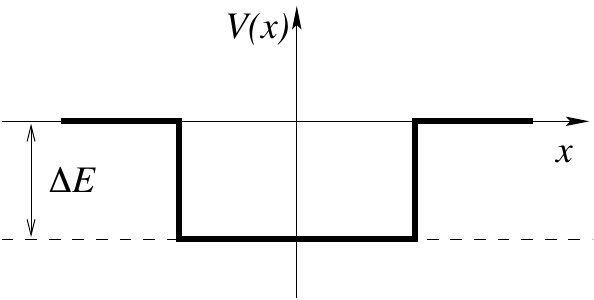}\end{center}\caption{Potential well}\label{fig:well}\end{figure}

One must be careful with this reasoning, since applied
carelessly it would lead to the same lifetime for the potential well
depicted in Figure \ref{fig:well} as for the potential plateau. (That is
because the reflection probability at an \emph{upward} potential step is,
as already mentioned, the same as that at a \emph{downward} potential step.) However, the potential well possesses bound states for
which the lifetime is infinite. In this regard it is important to bear
in mind that the symmetry in the reflection coefficient derived in \cite{LL}
always involves incoming waves at the same total energy $E>0$; for a
potential well it would thus say nothing about bound states, which have $E<0$.

This is a basic difference between confinement in a potential well and
paradoxical confinement on a potential plateau: In the well, the particle
has positive probability to stay forever. Mathematically speaking, the
potential well has bound states (i.e., eigenfunctions in the Hilbert space
$L^2(\mathbb{R})$ of square-integrable functions), whereas the potential
plateau does not. For the potential well, the initial wave packet will
typically be a superposition
$\psi = \psi_\mathrm{bound} +
\psi_\mathrm{scattering}$ of a bound state 
(a superposition of one or more 
square-integrable eigenfunctions) and a scattering state (orthogonal to all
bound states); then $\|\psi_\mathrm{bound}\|^2$ is the probability that the
particle remains in (a neighborhood of) the well forever. In contrast,
because of paradoxical reflection,  the
potential plateau has \emph{metastable states}, which remain on
the plateau for a long time but not forever---\y{namely, with lifetime \eqref{tqu}.}

\y{Let us give a cartoon of how one might expect these metastable states to behave, described in terms of the probability density function $\rho_t(x)$ at time $t$. Let $P_t$ denote the probability that the particle is in the plateau region at time $t$, $P_t=\int_{-a}^a \rho_t(x) \, dx$, and let us suppose that the particle is there initially, $P_0=1$. Assuming that the particle leaves the plateau at rate $\tau=\tau_\qu$, we have that $P_t=e^{-t/\tau}$. For simplicity, let us pretend that the distribution in the plateau region is flat, $\rho_t(x) = P_t/2a$ for $-a<x<a$. After leaving the plateau, the particle should move away from the plateau, say at speed $v$. Then $\rho_t(x)=0$ for $|x|>a+vt$ because such $x$ cannot be reached by time $t$, and the amount of probability between $x$ and $x+dx$ (with $a<x<a+vt$) at time $t$, $\rho_t(x)\, dx$, is what flowed off at $x=a$ between $\tilde{t}=t-(x-a)/v$ and $\tilde{t}-d\tilde{t} = t-(x+dx-a)/v$, which is half of the decrease in $P_t$ between $\tilde{t}-d\tilde{t}$ and $\tilde{t}$ (half because the other half was lost at $x=-a$), or}
\begin{equation}\label{rhomodel}
\rho_t(x)\, dx = \frac{1}{2} \biggl|\frac{dP_{\tilde{t}}}{d\tilde{t}}\biggr| d\tilde{t}=\frac{1}{2\tau}e^{-\tilde{t}/\tau}\, d\tilde{t}= \frac{1}{2v\tau} e^{-t/\tau}e^{(x-a)/v\tau} \, dx\,.
\end{equation}
\y{Likewise, for $-a-vt<x<-a$, $\rho_t(x)=  (1/2v\tau) e^{-t/\tau}e^{(|x|-a)/v\tau}$. This over-simplified model of $\rho_t(x)$ conveys a first idea of what kind of behavior to expect. Some of its features, notably the exponential increase with $|x|$ outside the plateau region, we will encounter again in the next sections.}

\y{In Section~\ref{sec:packetsonplateau} we investigate how a wave packet initially in the plateau region will behave. But before, in Section~\ref{sec:complexenergy}, we will compute the lifetime and confirm \eqref{tqu}. Our tool} will be a method similar to the stationary analysis of Section 2,
using special states lying outside the Hilbert
space $L^2(\RRR)$ (as do the stationary states of Section 2). And again
like the stationary states of Section 2, the special  states are
similar to eigenfunctions of the Hamiltonian: they are solutions
of the time-independent Schr\"odinger equation \eqref{TISE}, but with complex
``energy''!

\section{Eigenfunctions with ``Complex Energy''}
\label{sec:complexenergy}

We now derive the formula \eqref{tqu} for the lifetime $\tau =\tau_\qu$ from the behavior of solutions to the eigenvalue equation (\ref{TISE}), but with complex
eigenvalues. To avoid confusion, let us now call the eigenvalue $Z$ instead of $E$; thus, the equation reads
\begin{equation}\label{TISZ}
  Z \psi(x) = -\tfrac{\hbar^2}{2m} \psi''(x) + V(x) \, \psi(x)\,,
\end{equation}
where $V$ is the plateau potential as in \eqref{Vp}.
Such ``eigenfunctions
of complex energy'' were first considered by Gamow 
\cite{gamow1,laue} for the theoretical treatment of 
radioactive alpha decay.

The fact that the eigenvalue is complex may be confusing at first, since the Hamiltonian is a
self-adjoint operator, and it is a known fact that the eigenvalues of a self-adjoint operator are real. However, in the standard mathematical terminology for self-adjoint operators in Hilbert spaces, the words ``eigenvalue'' and ``eigenfunction'' are reserved for such solutions of \eqref{TISZ} that $\psi$ is square-integrable (= normalizable), i.e., $\psi \in L^2(\RRR)$. In this sense, all eigenvalues must be real indeed; for us this means that any solution $\psi$ of \eqref{TISZ} for $Z\in \CCC\setminus \RRR$ (where $\setminus$ denotes the set difference, i.e., \x{we require that} $\Im\, Z \neq 0$) is not square-integrable. In fact, even the eigenfunctions with real eigenvalue $E$ considered in \eqref{TISE} were not square-integrable, which means that they do not count as ``eigenfunctions'' in the mathematical terminology, and do not make the number $E$ an ``eigenvalue.'' Instead, $E$ is called an \emph{element of the spectrum} of the Hamiltonian. Still, the spectrum of any self-adjoint operator consists of real numbers, and thus $Z\in \CCC\setminus \RRR$ cannot belong to the spectrum of the Hamiltonian. So, the eigenvalues $Z$ we are talking about are neither eigenvalues in the standard sense, nor even elements of the spectrum. (Nevertheless we continue calling them ``eigenvalues,'' as they satisfy \eqref{TISZ} for some nonzero function.)

Let us explain how these complex eigenvalues can be useful in
describing the time evolution of wave functions.
Consider an eigenfunction $\psi$ with a complex eigenvalue $Z$. 
It generates a solution to the time-dependent Schr\"odinger equation
by defining
\be\label{eZ}
\psi(x,t) = \rme^{-\ci Zt/\hbar} \psi(x,0)\,.
\ee
The function grows
or shrinks exponentially with time, with rate given by the imaginary part
of $Z$. More precisely,
\be\label{psi2grow}
|\psi(x,t)|^2 = \rme^{2\Im\, Zt/\hbar} |\psi(x,0)|^2\,,
\ee
so that $2\Im\, Z/\hbar$ is the rate of growth of the density $|\psi(x,t)|^2$. For those eigenfunctions relevant to our purposes, the imaginary
part of $Z$ is always negative, so that $\psi$ shrinks with time. In particular, the amount of $|\psi|^2$ \x{in the high-potential region} decays with the exponential factor that occurs in \eqref{psi2grow}. Assuming that $|\psi|^2$ is proportional to the probability density at least in some region around the plateau (though not on the entire real line) for a sufficiently long time, and using that the lifetime $\tau$ for which the particle remains on the plateau is reciprocal to the decay rate of the amount of probability \x{in the plateau region}, we have that
\be\label{tauZ}
\tau = -\frac{\hbar}{2\Im\, Z}\,.
\ee
From \eqref{eZ} we can further read off that the phase of $\psi(x,t)$ at any fixed $x$ rotates with frequency $\Re\,Z/\hbar$, while for eigenfunctions with real eigenvalue $E$ it does so with frequency $E/\hbar$, which motivates us to call $\Re\,Z$ the energy and denote it by $E$. Thus,
\be
Z = E-\ci \frac{\hbar}{2\tau}\,.
\ee
Below, we will determine $\tau$ by determining the relevant eigenvalues $Z$, i.e., those corresponding to decay eigenfunctions, see below (\ref{Cbc}). 

The eigenfunctions $\psi$ differ from physical wave functions, among other respects, in that $|\psi|^2$ shrinks everywhere.  Since a local conservation law holds for $|\psi|^2$, this shrinking corresponds to a loss of $|\psi|^2$ at $x=\pm\infty$. What do these eigenfunctions have to do with physical wave functions? In the situation we want to consider, the physical wave function $\phi_t$ is such that the amount of $|\phi_t|^2$ \x{in the plateau region} continuously shrinks due to a flow of $|\phi|^2$ away from the plateau. On any large but finite interval $[-b,b]$ containing the plateau $[-a,a]$, $\phi_t$ may approach an eigenfunction $\psi_t$, and thus become a quasi-steady-state, i.e., stationary up to an exponential shrinking due to outward flux through $x=\pm b$ \y{(like the density $\rho_t(x)$ described around \eqref{rhomodel} for $t>(b-a)/v$). Indeed, this picture will be confirmed to some extent in Theorem~\ref{thm:metastable} below. It also suggests that, like $\rho_t$,} $\psi$ should grow exponentially in space as $x \to \pm \infty$: The density at
great distance from the plateau would be expected to agree with the
flow off the plateau in the distant past, which was
exponentially larger than in the present if the wave function in the 
plateau region shrinks exponentially with time. As we will see now in \eqref{eq:platansatz}--\eqref{Cbc}, the eigenfunctions do indeed grow exponentially with $|x|$ outside the plateau. 

\y{We now specify the eigenfunctions, starting with the general solution of \eqref{TISZ} without any requirements on the behavior at $\pm a$ (such as continuity of $\psi$ and $\psi'$). For $Z\in \CCC$ except $Z=0$ or $Z=-\Delta E$ it is}\footnote{For $Z=0$, the line for $-a<x<a$ has to be replaced by $A_0 +A_1x$; for $Z=-\Delta E$, $\psi(x) = D_- + E_- x$ for $x<-a$ and $\psi(x) = D_+ + E_+ x$ for $x>a$.}
\begin{align}\label{eq:platansatz}
\psi(x) = \begin{cases}
  B_- \rme^{-\ci \tilde{k} x}+C_- \rme^{\ci \tilde{k} x} & \text{when } x<-a, \\
  A_+ \rme^{\ci k x}+A_- \rme^{-\ci k x} & \text{when } -a<x<a, \\
  B_+ \rme^{\ci \tilde{k} x}+C_+ \rme^{-\ci \tilde{k} x} & \text{when } x>a,
 \end{cases}
\end{align}
where
\be\label{kdef}
k=\sqrt{2mZ}/\hbar \quad \text{and} \quad 
\tilde{k} =\sqrt{2m(Z+\Delta E)}/\hbar
\ee
with the following (usual) definition of the complex square
root:
            Given a complex number $\zeta$ other than one that is real
            and $\leq 0$, let $\sqrt{\zeta}$ denote the square root with
            positive real part, $\Re\,\sqrt{\zeta}>0$.  For $\zeta \le 0$, we
            let $\sqrt{\zeta}=\ci \sqrt{|\zeta|}$.  (Since
            (\ref{eq:platansatz}) remains invariant under changes in the
            signs of $k$ and $\tilde{k}$, choosing the positive branch for
            the square roots is not a restriction for the solutions.) 

\y{We remind the reader that a term like $B_+ \rme^{\ci \tilde{k} x}$ is \emph{not} a plane wave since $\tilde{k}$ is not real but complex. It is the product of a plane wave and an exponential growth factor governed by the imaginary part of $\tilde{k}$.}
            
            We
            are interested only in those solutions for $Z$ with $\Re\, Z = E
            > 0$; these are the ones that should be relevant to the
            behavior of states starting out on the plateau (with positive
            energy). 
Nevertheless, for mathematical simplicity, we will also allow $\Re\, Z \leq 0$ but exclude any $Z$ that is real and negative or zero. For any $Z\in \CCC\setminus(-\infty,0]$ we have that $\Re\,\tilde{k} > 0$, so that the probability current $j$ associated with $\exp(\ci\tilde{k}x)$ is positive (i.e., a vector pointing to the right), namely $j= (\hbar/m)|\psi|^2\Re\,\tilde{k}$. Since we do not want to consider any contribution with a current from infinity to the plateau, we assume that
\be\label{Cbc}
C_+ = C_-=0\,.
\ee
\y{Thus, the kind of eigenfunction relevant to us is what we define to be} a \emph{decay eigenfunction} or \emph{Gamow eigenfunction}: a nonzero function $\psi$ of the form \eqref{eq:platansatz} with \eqref{kdef} and $C_\pm=0$, satisfying the eigenvalue equation \eqref{TISZ} except at $x=\pm a$ (where $\psi''$ does not exist) for some $Z\in\CCC\setminus (-\infty,0]$, such that both $\psi$ and $\psi'$ are continuous at $\pm a$; those $Z$ that possess a decay eigenfunction we call \emph{decay eigenvalues} or \emph{Gamow eigenvalues}.\footnote{Here is a look at the negative $Z$'s that we excluded in this definition: In fact, for $Z\in (-\infty,0] \setminus \{-\Delta E\}$ there exist no nonzero functions with $C_\pm =0$ satisfying the eigenvalue equation \eqref{TISZ} for all $x\neq\pm a$ such that $\psi$ and $\psi'$ are continuous. (But we have not included the proof in this paper.) For $Z=-\Delta E$, the coefficients $C_\pm$ are not defined, so the condition \eqref{Cbc} makes no sense.}

The remaining coefficients $A_\pm,B_\pm$, \y{as well as the possible values of $Z$, $k,\tilde{k}$ are determined (up to an overall factor for $A_\pm,B_\pm$) from \eqref{eq:platansatz}, \eqref{kdef}, \eqref{Cbc}} by the requirement that both $\psi$ and its derivative
$\psi'$ be continuous at $\pm a$, the ends of the plateau. 
We have collected the details of the computations into Appendix 
\ref{sec:plateigen}, and report here the results. \y{To express them, we use the natural unit of energy in this setting, which is the energy whose de Broglie wavelength is equal to the length $2a$ of the plateau,}
\begin{equation}\label{Wdef}
W := \frac{\pi^2 \hbar^2}{2ma^2}.
\end{equation}
\y{For paradoxical confinement to occur, $\Delta E$ should be large compared to $W$. The eigenvalues $Z$ relevant to paradoxical confinement are those whose real part, the energy $\Re \, Z = E$, is positive (because we want to look at states starting on top of the plateau) and small (because only states of small energy are affected by paradoxical reflection), and whose imaginary part, $\Im\, Z = -\hbar/2\tau$, is negative (because eigenfunctions with $\Im\, Z>0$ would grow with $t$, rather than shrink, due to influx from $x=\pm\infty$) and small (because they have large lifetime $\tau$). In particular, we are not interested in eigenvalues $Z$ far away from zero.}


\begin{Thm}\label{thm:eigenvalues}
Suppose $\Delta E \ge 100 W$. Then the number $N$ of decay eigenvalues $Z$ of \eqref{TISZ} with $|Z|\leq \Delta E/4$ lies in the range $\sqrt{\Delta E/W}-2<N\le \sqrt{\Delta E/W}+2$. There is a natural way of numbering these eigenvalues as $Z_1,\ldots,Z_N$. (There is no formula for the eigenvalue $Z_n$, but it can be defined implicitly.) With each $Z_n$ is associated a unique (up to a factor) eigenfunction $\psi_n$, and $|\psi_n(x)|$ is exponentially increasing as $x\to \pm \infty$ \y{(i.e., $\Im\, \tilde{k}<0$ for $\psi_n$)}. Furthermore, for $n\ll \sqrt{\Delta E/W}$, 
\be\label{Zn}
Z_n \approx  \biggl( \frac{W}{4}-\ci \frac{W^{3/2}}{2\pi \sqrt{\Delta E}}\biggr) n^2\,.
\ee 
\end{Thm}

\begin{figure}[ht]\begin{center}\includegraphics[width= \textwidth]{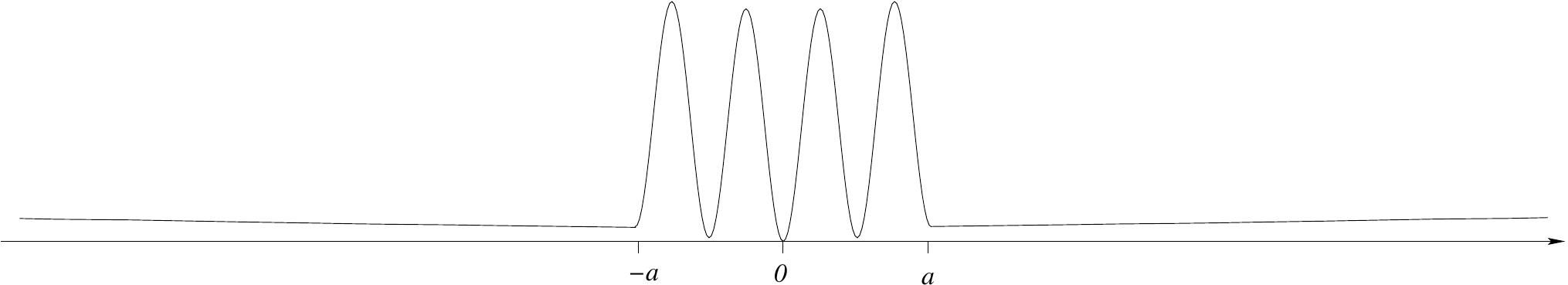}\end{center}\caption{Plot of $|\psi_n(x)|^2$ for an eigenfunction $\psi_n$ with complex eigenvalue according to \eqref{TISZ} with $V(x)$ the plateau potential as in Figure~\ref{fig:plateau}; the parameters are $n=4$ and $\Delta E = 64 \, W$; in units with $a=1$, $m=1$, and $\hbar=1$, this corresponds to $\Delta E=32\pi^2 = 315.8$.}\label{fig:eig}\end{figure}

The proof is given in Appendix~\ref{sec:plateigen}; to our knowledge, the values \eqref{Zn} are not in the literature so far. 
\y{The precise meaning of approximate equalities $x\approx y$ is $\lim x/y=1$ as $\Delta E\to\infty$ and $a,n$ are fixed.} 

What can we read off about the lifetime $\tau$? 
In the regime $\Delta E \gg W$, the $n$-th complex eigenvalue $Z_n$ with $n\ll \sqrt{\Delta E/W}$ is such that
\begin{equation}\label{complexE}
  \Re \, Z_n \approx \frac{\hbar^2 \pi^2 n^2}{8 m a^2} \,,\quad
  \Im \, Z_n \approx - \frac{2\hbar}{a \sqrt{2m \Delta E}}\Re \, Z_n \,.
\end{equation}
(Readers familiar with the infinite well
potential, $V(x)=0$ when $-a \le x \le a$ and $V(x) = \infty$ when $|x|>a$, which corresponds to
the limit $\Delta E \to \infty$ of very deep wells of the type shown in Figure~\ref{fig:well}, will notice that $\Re \, Z_n$ given above
actually coincides
with the eigenvalues of the infinite well potential of length $2 a$.) 
Using \eqref{tauZ} and $E=\Re\,Z$, one finds that
\begin{equation}\label{tasymp}
  \tau \approx a\frac{\sqrt{2m\Delta E}}{4 E} =
 \frac14\sqrt{\frac{\Delta E}{E}}\, \tau_\cl = \tau_\qu \,,
\end{equation} 
the same value as specified in \eqref{tqu}. This completes our derivation of the
lifetime \eqref{tqu} from complex eigenvalues. 

As we did for the potential step,
we now also look at the question whether wave packets behave in the same way as the eigenfunctions, that is, whether a wave packet can remain in the plateau region for the
time span \eqref{tqu}.

\section{Wave Packets on the Plateau}
\label{sec:packetsonplateau}

We will now use the eigenfunctions to draw conclusions about the behavior of normalized (square-integrable) wave packets (see \cite{DGK10} for similar considerations about radioactive decay). We will first show that, for large $\Delta E$, there actually exist normalized wave packets, initially concentrated in the plateau interval and leaking out at an exponential, but slow, rate.  The wave packets approach a quasi-steady-state situation in an expanding region surrounding the plateau---one that differs from a genuine steady state in that there is a global uniform overall exponential decay in time. This picture is very similar to the behavior of the model $\rho_t(x)$ described around \eqref{rhomodel}. Within the expanding region, the wave function is approximately given by an eigenfunction with complex eigenvalue as described in the previous section. An explicit example of a normalized wave packet that behaves in this way is given by cutting off an eigenfunction outside the plateau interval. This is the essence of Theorem~\ref{thm:metastable} below. These results can partially be generalized for other compactly supported potentials, not just for the plateau potential considered here. This however requires more advanced mathematical tools: see \cite{skibsted86,CH09,DGK10}.

%

\y{We write $\psi_{n}$ for the eigenfunction with eigenvalue $Z_n$, $\psi_{n,t}=\rme^{-iZ_nt/\hbar}\psi_n$ for the time-evolved eigenfunction, $\tau_n = -\hbar/2\Im\,Z_n$ for the corresponding decay time, and}
\be\label{vdef}
v_n = \frac{\hbar}{m} \Re \, \tilde{k}_n\,
\ee
\y{for} the speed at which an escaping particle moves away
from the plateau.

\begin{Thm}\label{thm:metastable}
Let $n$ be a fixed positive integer; keep \y{the plateau length $2a$ fixed and consider the regime $\Delta E \gg W$.} The initial wave function
\be\label{varphi0}
\varphi_0 (x) = \begin{cases} 
A_n \psi_n(x) & \text{for }-a\leq x\leq a\\
0 & \text{otherwise,}\end{cases}
\ee
with normalization constant $A_n$, evolves with time in such a way that, for every $0<t<\tau_n$, $\varphi_t$ is close to $A_n\psi_{n,t}$ on the interval $[-a-v_nt,a+v_nt]$ growing at speed $v_n$. Explicitly, for $0<t<\tau_n$,
\be\label{closeavt}
\int_{-a-v_nt}^{a+v_nt} \bigl|\varphi_t(x) - A_n \psi_{n,t}(x) \bigr|^2 \rmd x \ll 1\,.
\ee
\end{Thm}

The proof is included in Appendix \ref{sec:metastable}. \y{Theorem~\ref{thm:metastable} is used in the proof of Theorem~\ref{thm:confinement1}.}

\y{Theorem~\ref{thm:metastable} provides a deeper justification of the formula \eqref{tqu} for the decay time $\tau_\qu$ by showing that $1/\tau_\qu$ is not merely the decay rate of \emph{eigenfunctions} $\psi_n$ but also the decay rate of certain \emph{normalized wave packets} $\varphi_t$. The amount of probability in $[-a,a]$ indeed shrinks at rate $1/\tau_\qu=1/\tau_n$, at least up to time $\tau_n$. In particular, the particle has probability $\approx 1/\rme=0.3679$ to stay in the plateau region until $\tau_n$.}

\y{It may seem that Theorem~\ref{thm:metastable} concerns only a specially chosen wave packet $\varphi_0$, but by forming linear combinations we can obtain the slow decay for any wave packet of low energy:}

\begin{Corol}\label{corr:metastable}
\y{Let $\Delta E \gg W$.} For any initial wave function $\psi$ on the plateau with contributions only from eigenfunctions $\psi_n$ with low $n$, i.e.,
\begin{align}\label{comb}
\psi(x) = 
\begin{cases} 
\displaystyle \sum_{n=1}^{n_\mathrm{max}} c_n \, \psi_n(x) & \text{when } -a\le x \le a,\\
0 & \text{otherwise,}
\end{cases}
\end{align}
with $\Delta E$-independent $n_{\max}$ and coefficients $c_n$, the time-evolved
wave function $\psi_t = \rme^{-\ci Ht/\hbar} \psi$ is close to $\sum_n c_n\, \psi_{n,t}$ on the interval $[-a-vt,a+vt]$ growing at speed $v=\min(v_1,\ldots,v_{n_\mathrm{max}})$, at least up to time $\min (\tau_1, \ldots, \tau_{n_{\max}})$. That is, 
\be\label{combcomp}
\int_{-a-vt}^{a+vt} \bigl| \psi_t(x) - \sum_n c_n \, \psi_{n,t}(x) \bigr|^2 \rmd x 
\ll 1
\ee
for $t\leq\min (\tau_1, \ldots,\tau_{n_{\max}})$.
\end{Corol}

\bigskip

This means that any such wave packet $\psi$ will have a long decay time on the plateau, namely at least $\min (\tau_1,\ldots, \tau_{n_{\max}})$ (with each $\tau_n$ given by the quantum formula \eqref{tqu} and not by the classical formula \eqref{tcl}!); indeed, \eqref{combcomp} suggests that the decay time of $\psi$ is of the order of the largest $\tau_n$ with $1\leq n \leq{n_{\max}}$ and significant $|c_n|^2$. 

As a final remark we note that the decay results described here, both
qualitative and quantitative, presumably apply as well to the standard
tunnelling situation in which a particle is confined inside a region by a
potential barrier (a wall) that is high but not infinitely high, separating
the inside from the outside. For this situation, more detailed results were
obtained in \cite{CC99} by other methods based on analytic continuation.

\section{Conclusions}

We have argued that paradoxical reflection, the phenomenon that a quantum particle can be reflected at a sudden drop in the potential, and paradoxical confinement, the phenomenon that a quantum particle tends to remain in a potential plateau region, are real phenomena and
not artifacts of the stationary analysis. We have pointed out that the
effect is a robust prediction of the Schr\"odinger equation, as it persists
when the potential step is not assumed to be rectangular but soft, and when
the incoming wave is a packet of finite width. We have provided numerical
evidence and identified the relevant
conditions on the parameters. We have explained why it is not a 
counter-argument to note that paradoxical reflection is classically
impossible. We conclude that paradoxical reflection is a fact, not an artifact.
Finally, we have computed that a state (of sufficiently low 
energy) on a potential plateau as in Figure~\ref{fig:plateau} has a long 
decay time, no less than $\tau_\qu$ given by \eqref{tqu}. We conclude from this 
that a plateau potential can, for suitable parameters, effectively be
confining. Thus, the effect could indeed be used for constructing a
(metastable) particle trap.

\appendix

\section{Solving the Plateau Eigenvalue Equation}\label{sec:plateigen}

We now prove Theorem~\ref{thm:eigenvalues}; that is, we determine all decay eigenfunctions of \eqref{TISZ}, as defined after \eqref{Cbc}. The continuity of $\psi$ requires that
\begin{align}
 A_+ \rme^{\ci k a} +  A_- \rme^{-\ci k a} & = B_+ \rme^{\ci \tilde{k} a}\,, \\
 A_+ \rme^{-\ci k a} +  A_- \rme^{\ci k a} & = B_- \rme^{\ci \tilde{k} a}\,, 
\end{align} 
and continuity of $\psi'$ that also
\begin{align}
 k \left( A_+ \rme^{\ci k a} - A_- \rme^{-\ci k a}\right) 
 & = \tilde{k} B_+ \rme^{\ci \tilde{k} a}, \\
  k \left( A_+ \rme^{-\ci k a} -  A_- \rme^{\ci k a}\right) 
  & = -\tilde{k} B_- \rme^{\ci \tilde{k} a}. 
\end{align} 
Recall that both $k$ and $\tilde{k}$ can be complex.
Since we assume $\Delta E> 0$, and since, by \eqref{TISZ},
$\tilde{k}^2=k^2+2m\Delta E/\hbar^2,$ we have that $k\pm \tilde{k} \ne
0$, and these equations are readily solved.  First, we find the relations
\begin{align}
A_- & = \rme^{\ci 2 a k} \frac{k-\tilde{k}}{k+\tilde{k}}  A_+,\label{eq:A}\\
B_+ & = \rme^{\ci a (k-\tilde{k})} \frac{2 k}{k+\tilde{k}} A_+, \label{eq:Bplus}\\
B_- & = \rme^{-\ci a (k+\tilde{k})} \frac{2 k}{k-\tilde{k}} A_+, \label{eq:Bminus}
\end{align}
with the additional requirement that, since  $A_+\ne 0$ for decay eigenfunctions,
\begin{align}\label{eq:kkpeq}
 \left(\frac{k+\tilde{k}}{k-\tilde{k}}\right)^2 = \rme^{\ci 4 a k} .
\end{align}
\y{Let}
\begin{align}\label{eq:defld}
  \lambda_0 = \frac{2 \pi \hbar}{\sqrt{2 m \Delta E}},\qquad
  \alpha = \frac{a}{\pi \hbar}\sqrt{2 m \Delta E} = \frac{2 a}{\lambda_0}.
\end{align}
\y{$\lambda_0$ is the de Broglie wavelength corresponding to the height $\Delta E$ of the
potential plateau, and $\alpha$ is the width of the plateau in units of $\lambda_0$. Thus in terms of $W$ defined in (\ref{Wdef}) we have $\alpha=\sqrt{\Delta E/W}$.
In} order to express $k$ in natural units, let
\be\label{wdef}
\kappa := \frac{\lambda_0 k}{2 \pi}\,.
\ee
Then
\be\label{tildekw}
k=\frac{2\pi}{\lambda_0}\kappa\,, \qquad
\tilde{k} = \frac{2 \pi}{\lambda_0}\sqrt{1+\kappa^2}\,,
\ee
and we have that
\begin{align}
\frac{k+\tilde{k}}{k-\tilde{k}} = 
\frac{\kappa+\sqrt{1+\kappa^2}}{\kappa-\sqrt{1+\kappa^2}} = 
-\left(\kappa+\sqrt{1+\kappa^2}\right)^2.
\end{align}
Thus (\ref{eq:kkpeq}) is equivalent to
\begin{align}\label{eq:kkpeq2}
 \left(\kappa + \sqrt{1+\kappa^2}\right)^4 = \rme^{\ci 4 \pi \kappa \alpha} .
\end{align}
The solutions of this equation coincide with those of the equation
\begin{align}\label{eq:lnweq}
 \ln \left(\kappa + \sqrt{1+\kappa^2}\right) = \ci \pi \kappa \alpha - \ci \frac{\pi n}{2}
\end{align}
where $n\in \ZZZ$ is arbitrary and $\ln $ denotes the principal branch of the complex logarithm.\footnote{For $\zeta\in \CCC\setminus \{0\}$, the equation $\rme^z = \zeta$ has infinitely many solutions $z$, all of which have real part $\ln |\zeta|$, and the imaginary parts of which differ by integer multiples of $2\pi$; by $\ln \zeta$ we denote that $z$ which has $-\pi<\Im \,z \le \pi$.} 

Thus, with every decay eigenfunction $\psi$ is associated a solution $\kappa$ of \eqref{eq:lnweq} (with $\Re\, \kappa>0$, since $\Re\, k>0$ by definition \y{\eqref{kdef}} of $k$) and an integer $n$. Furthermore, $n\ge -1$, because $\Re \, \kappa>0$ and the imaginary part of the left hand side of \eqref{eq:lnweq} must lie between $-\pi$ and $\pi$. Conversely, with every solution $\kappa$ of \eqref{eq:lnweq} with $\Re\, \kappa>0$ there is associated a decay eigenvalue
\be\label{Zw}
Z= \kappa^2 \Delta E 
\ee
and an eigenfunction $\psi$ that is unique up to a factor: Indeed, \eqref{wdef} and \eqref{tildekw} provide the values of $k$ and $\tilde{k}$ and imply \eqref{Zw} and \eqref{eq:kkpeq}; $\Re\, \kappa>0$ implies $Z\notin(-\infty,0]$, as well as $\Re\, k>0$, so that indeed $k=\sqrt{2mZ}/\hbar$; $k\pm\tilde{k} \neq 0$; $A_+$ can be chosen arbitrarily in $\CCC\setminus\{0\}$, and if $A_-$ and $B_\pm$ are chosen according to \eqref{eq:A}--\eqref{eq:Bminus} then $\psi$ is nonzero (as, e.g., $B_+\neq 0$ when $k\neq 0$ and $A_+\neq 0$) and a decay eigenfunction. We note that the condition $\Re \, \kappa>0$ is automatically satisfied when $n\ge 2$, as we can read off from the imaginary part of \eqref{eq:lnweq} using that $\ln$ has imaginary part in $(-\pi,\pi]$. 

To determine $\psi$ explicitly, note that
$\rme^{\ci 2 a k} \frac{k-\tilde{k}}{k+\tilde{k}} = (-1)^{n+1}$, and thus
$A_- = (-1)^{n+1} A_+$, $B_- = (-1)^{n+1}B_+$; setting $A_+=\frac{1}{2}$ and introducing the notation
\begin{align}
B:=B_+\rme^{\ci a \tilde{k}}=\rme^{\ci a k} \frac{k}{k+\tilde{k}}=
\rme^{\ci \pi \kappa \alpha} \frac{\kappa}{\sqrt{1+\kappa^2}+\kappa}=\ci^n \kappa\,,
\end{align}
we obtain that for odd $n$
\begin{align}\label{eq:platcosres}
& \psi(x) = B
 \left[\chi(x>a) \rme^{\ci \tilde{k} (x-a)}+ \chi(x<-a) \rme^{-\ci \tilde{k}
   (x+a)}\right]
+ \chi(-a\le x\le a) \cos \left(k x\right),
\end{align}
and for even $n$
\begin{align}\label{eq:platsinres}
& \psi(x) = B
 \left[\chi(x>a) \rme^{\ci \tilde{k} (x-a)}- \chi(x<-a) \rme^{-\ci
     \tilde{k} (x+a)}\right]
+ \chi(-a\le x\le a) \sin \left(k x\right)
\end{align}
with the notation 
$\chi(Q)$ to denote the characteristic function of a condition $Q$:
\begin{align}\label{chidef}
  \chi(Q) = \begin{cases}
  1 & \text{when } Q \text{ is true}, \\
  0 & \text{otherwise}.
 \end{cases}
\end{align}

To sum up what we have so far, the decay eigenvalues are characterized, via \eqref{Zw}, through the solutions $\kappa$ of \eqref{eq:lnweq} with $\Re\, \kappa>0$. In order to study existence, uniqueness, and the asymptotics for $\alpha\to \infty$ of these solutions, let us now assume, as in Theorem~\ref{thm:eigenvalues}, that $\alpha \ge 10$ and $|Z| \le \Delta E/4$. By virtue of \eqref{Zw}, the latter assumption is equivalent to $|\kappa|\le 1/2$. We first show that solutions with $|\kappa|\le 1/2$ must have $|n|\le \alpha+2$: Since $\ln$ has imaginary part in $(-\pi,\pi]$, \eqref{eq:lnweq} implies that $\Re\, \kappa \in (\tfrac{n-2}{2\alpha},\tfrac{n+2}{2\alpha}]$, and hence
\be
\frac{1}{2}\geq |\kappa| \geq |\Re\,\kappa| \geq \frac{|n|-2}{2\alpha}\,,
\ee
or $|n|\leq \alpha+2$. Next recall that for decay eigenvalues, $n\ge -1$, so we obtain at this stage that the number of values that $n$ can assume is at most $\alpha+4$, as the possible values are $-1,0,1,2,\ldots\le \alpha+2$. We will later exclude $n=0$ and $n=-1$.

We now show that there exists a unique solution $\kappa$ of \eqref{eq:lnweq} for every $n$ with $|n|\leq \alpha+2$.  Let
\be
F(\kappa)=\frac{n}{2 \alpha} -\frac{\ci}{\pi \alpha} 
\ln \left(\kappa + \sqrt{1+\kappa^2}\right)\,,
\ee
so that \eqref{eq:lnweq} can equivalently be rewritten as the fixed point equation
\be
F(\kappa) = \kappa\,.
\ee
We use the Banach fixed point theorem \cite{Ban} to conclude the existence and uniqueness of $\kappa$. Since
\begin{align}\label{eq:F'}
  F'(\kappa) = -\frac{\ci}{\pi \alpha} \frac{1}{\sqrt{1+\kappa^2}}\,,
\end{align}
we have, by the triangle inequality, that
\be
|F'(\kappa)|= \frac{1}{\pi \alpha |1+\kappa^2|^{1/2}}\le \frac{1}{\pi \alpha |1-|\kappa|^2|^{1/2}}\,.
\ee
Let us consider for a moment, instead of $|\kappa|\le 1/2$, the disk $|\kappa|\le r$ for any radius $0<r<\sqrt{1-1/\pi^2 \alpha^2}$. There we have that $|F'(\kappa)|\le 1/(\pi \alpha \sqrt{1-r^2})=:K <1$. Thus, for any $\kappa,\kappa'$ in the closed disk of radius $r$, $|F(\kappa')-F(\kappa)|\le K  |\kappa'-\kappa|$, and, using $|F(0)|=\frac{|n|}{2 \alpha}$,
\begin{align}\label{eq:ball}
  |F(\kappa)|\le |F(\kappa)-F(0)|+|F(0)|\le rK +
  \frac{|n|}{2 \alpha}\le r\,,
\end{align}
provided that
\be\label{contractcond}
|n|\le 2\alpha r(1-K )\,.
\ee
Thus, in this case, $F$ is a contraction in the ball of radius $r$, with a contraction
constant of at most $K$.  By the Banach fixed point
theorem there is then a unique solution to the equation $F(\kappa)=\kappa$ in the
ball $|\kappa|\le r$. Even though we are ultimately interested in the radius $1/2$, let us set $r=1/\sqrt{2}$, which satisfies $r<\sqrt{1-1/\pi^2\alpha^2}$ as $\alpha\ge 10$; also \eqref{contractcond} is satisfied because $|n|\le \alpha+2$ and $\alpha \ge 10> 2(1+1/\pi)/(\sqrt{2}-1)\approx 6.37$. Hence, for every $n$ with $|n|\le \alpha+2$, there is a unique solution $\kappa_n$ with $|\kappa_n|\le 1/\sqrt{2}$.

Getting back to the ball of radius $1/2$, while some of the $\kappa_n$ may have modulus greater than $1/2$, we can at least conclude that there is at most one solution with modulus $\le 1/2$ for every $n$ with $|n|\le \alpha+2$. In addition, by setting $r=1/2$, we obtain from \eqref{contractcond} that $|\kappa_n|\le 1/2$ for every $n$ with $|n|\le \alpha-1$.
If $n=0$, then $F(0)=0$ and $\kappa_0=0$ is the unique solution, which would lead to $\psi=0$. This excludes $n=0$. Which of the solutions have $\Re\,\kappa_n>0$, as required for decay eigenvalues? 
For any $n$ with $|n|\le \alpha+2$, let $\kappa_n^{(j)}$ be defined recursively by 
$\kappa_n^{(j+1)}=F(\kappa_n^{(j)})$ with $\kappa_n^{(0)}=0$.  
Then, again by the Banach fixed point theorem for $r=1/\sqrt{2}$, 
$\kappa_n^{(j)} \to \kappa_n$ as $j\to \infty$, and
\begin{align}\label{eq:wnest}
 |\kappa_n-\kappa_n^{(j)}|\le \frac{K^j}{1-K}|\kappa_n^{(1)}-\kappa_n^{(0)}|
\le |n| \alpha^{-(j+1)} \,.
\end{align}
For $n=-1$ and $j=1$, this gives us that $|\kappa_{-1}-\kappa_{-1}^{(1)}|\le \alpha^{-2}$, and with $\kappa_{-1}^{(1)}=-1/2\alpha$ and $\alpha\ge 10$, we can conclude that $\Re\,\kappa_{-1}<0$. This excludes $n=-1$.  For $n>0$, in contrast, the fact that $|\kappa_n-\kappa_n^{(1)}|\le |n|\alpha^{-2}$ allows us to conclude, with $\kappa_n^{(1)}=n/2\alpha$ and $\alpha\ge 10$, that $\Re\, \kappa_n>0$. Hence, the decay eigenvalues with $|Z|\le \Delta E/4$ are in one-to-one correspondence with those $\kappa_n$, $0<n\le \alpha+2$, that have $|\kappa_n|\le 1/2$; the number of these $\kappa_n$ must, as we have shown, be greater than $\alpha-2$ and less than or equal to $\alpha+2$. 

\xj{
Furthermore, for these $\kappa_n$, $\Im\, \kappa_n<0$: Computing $\kappa_n^{(2)}$ explicitly yields
\begin{align}\label{eq:wn2expl}
  \kappa_n^{(2)} = \nu -\ci \frac{1}{\pi \alpha}\ln \left(\nu + \sqrt{1+\nu^2}\right) 
  \qquad\text{with }\nu=\frac{n}{2 \alpha}.
\end{align}
Using \eqref{eq:wnest} as before, the claim follows if we can show that $\Im \, \kappa_n^{(2)}<-n\alpha^{-3}$.
We claim that for all $x\ge 0$,
\be\label{claim}
\frac{1}{\sqrt{1+x^2}}x \le \ln(x+\sqrt{1+x^2}) \le x\, .
\ee
Since $0<\nu\le (\alpha+2)/2\alpha \le 0.6$ by the assumption $\alpha\ge 10$, for such $\nu$ and $\alpha$ we then have
$-\Im \, \kappa_n^{(2)}\ge \frac{1}{\sqrt{2} \pi \alpha}\nu>n\alpha^{-3}$, and thus $\Im\, \kappa_n<0$.
The inequalities \eqref{claim} can be derived as follows. Consider the function $f(x)=\ln(x+\sqrt{1+x^2})-x$,
for which $f(0)=0$ and $f'(x)=\frac{1}{\sqrt{1+x^2}}-1$.  Thus $f(x)=\int_0^x\!\rmd y\,f'(y)$, and
$-1+\frac{1}{\sqrt{1+x^2}}\le f'(y) \le 0$ for all $0\le y\le x$, which immediately yields the bounds in \eqref{claim}. 
}

\xj{
As a consequence of $\Im\,\kappa_n<0$ (and $\Re\,\kappa_n>0$), also $\Im\, \tilde{k}<0$, so that $|\psi(x)|$ grows exponentially as $x\to \pm\infty$.  By $|\kappa_n-\kappa_n^{(2)}|\le n/\alpha^3$ and the above estimates for 
$\Im \, \kappa_n^{(2)}$, we also have the following explicit bounds for the real and imaginary parts of $\kappa_n$,
\begin{align}\label{eq:reexplbound}
\frac{n}{2\alpha}\Bigl(1-\frac{2}{\alpha^2}\Bigr) & \le \Re\, \kappa_n
\le \frac{n}{2\alpha}\Bigl(1+\frac{2}{\alpha^2}\Bigr)\, ,
\\ 
\frac{n}{2\pi \alpha^2}\Bigl(1-\frac{2\pi}{\alpha}-\frac{n^2}{4\alpha^2}\Bigr) &\le -\Im\, \kappa_n
\le \frac{n}{2\pi \alpha^2}\Bigl(1+\frac{2\pi}{\alpha}\Bigr)\, , \label{eq:imexplbound} 
\end{align} 
where in the second formula, we have simplified the result using the bound 
$1/\sqrt{1+\nu^2}\ge 1/(1+\nu^2) \ge 1-\nu^2$.
}

Now let us consider the asymptotics for $n\ll \alpha$. From \eqref{eq:wnest} we have that
$\kappa_n$ is given by the right hand side of \eqref{eq:wn2expl} up to an error of order $O(n \alpha^{-3})$. 
Therefore, for integers $n$ with $0<n \ll \alpha$ we have that
\begin{align}\label{eq:smallnest} 
  k_n \approx \frac{\pi n}{2a}-\ci \frac{n}{2a \alpha}\,,
\quad
  \tilde{k}_n \approx \frac{\pi\alpha}{a} -\ci \frac{n^2}{4a\alpha^2}\,,
\quad
  Z_n = \kappa_n^2 \Delta E 
 \approx \frac{ n^2\Delta E}{4 \alpha^2} \left( 1- \ci \frac{2}{\pi \alpha}\right)\,.
\end{align}
\xj{
The previous estimates, in particular
inequalities \eqref{eq:reexplbound} and \eqref{eq:imexplbound}, can be used to estimate the accuracy of these approximations. For instance,
\begin{align}\label{eq:reknest}
 \Bigl|k_n-\frac{\pi n}{2a}\Bigr|=\frac{2 \pi}{\lambda_0}|\kappa_n-\kappa_n^{(1)}|
  \le \frac{2 \pi}{\lambda_0} \frac{n}{\alpha^2} =  \frac{\pi n}{a\alpha}\, .
\end{align}
Also, since $-\Im\, Z_n= 2 \Delta E\, \Re\, \kappa_n (-\Im \kappa_n)$, the lifetimes $\tau_n$ satisfy
\begin{align}
 C_1 \frac{\pi \alpha^3 \hbar}{n^2 \Delta E}\le \tau_n \le 
 C_2 \frac{\pi \alpha^3 \hbar}{n^2 \Delta E}\, ,
\end{align}
for all $0<n\le \alpha$, and with some numerical constants $C_1,C_2>0$.  Using the definition of $\alpha$, here $\pi \alpha^3 \hbar/(n^2 \Delta E)=\frac{m a^2}{\pi\hbar}\alpha n^{-2}$.
Thus if we consider the limit $\Delta E\to \infty$ while keeping all other parameters fixed,
we have $\alpha\to \infty$ and can choose
$C_1=1-O(\alpha^{-1})$ and $C_2=1+O(\alpha^{-1})$. Therefore, $\tau_n/\alpha \to \frac{m a^2}{\pi\hbar} n^{-2}$ for any fixed $n$; in particular, $\tau_n\to \infty$.
}

\section{Derivation of the Lifetime Estimates for the Meta\-stable States in the Plateau Region}
\label{sec:metastable}

\noindent\textit{Proof of Theorem \ref{thm:metastable}.} 
We construct an auxiliary function $f(x,t)$ which does not obey the Schr\"odinger equation but remains close to the time-evolved eigenfunction in a growing region around the plateau.  We then prove that this function forms an excellent approximation of $\varphi_{t}(x)$. We will define $f(x,t)$ by cutting off the time-evolved eigenfunction $\rme^{-\ci t Z_n/\hbar} \psi_n$, though in a continuous way using Gaussians with time-dependent parameters.

\xj{
We begin by estimating the normalization constant $A_n$.  For this, we define for all $|x|\le a$ and integers $n\ge 1$,
$\phi_n(x)=\cos(\frac{\pi n}{2 a} x)$, if $n\ge 1$ is odd, and
$\phi_n(x)=\sin(\frac{\pi n}{2 a} x)$, if $n\ge 2$ is even.  A short computation shows that 
$\phi_n(x) = \pm \sin(\frac{\pi n}{2 a} (x+a))$, and thus the collection of functions $(\phi_n)$ is up to a constant equal to the sine-basis of 
square integrable functions on $[-a,a]$.  We also define $\phi_n(x)=0$ for $|x|>a$.  Since $\int_{-a}^a |\phi_n(x)|^2=a$,
their normalization constants are independent of $n$, all equal to $a^{-1/2}$.  
By \eqref{eq:reknest} for any $n$ the difference $z_n=k_n-\frac{\pi n}{2a}$ satisfies $|z_n|\le \frac{\pi n}{a\alpha}$.
Therefore, by expanding the appropriate cosine or sine, we find for all $|x|\le a$, 
\begin{align}
 \bigl|\psi_n(x)-\phi_n(x)\bigr|\le |1-\cos(z_n x)|+|\sin(z_n x)|\le (|z_n x|^2 +|z_n x|) \rme^{|z_n x|}\, ,
\end{align}
where $|z_n x|\le \pi n/\alpha\le\pi (1+2/\alpha)$.  Thus there is a pure constant $c$ such that 
\begin{align}
 \int_{-a}^a\bigl|\psi_n(x)-\phi_n(x)\bigr|^2\le a c^2 \frac{n^2}{\alpha^2}\, .
\end{align}
($c=2 \pi$ will suffice, if $n/\alpha$ is small enough.)
By the triangle inequality and the definition of the normalization constant $A_n>0$, the left hand side has a lower bound
$|A_n^{-1}-\sqrt{a}|^2$.  Thus $A_n = a^{-1/2}+ O(n/\alpha)$, and already if $n \le \alpha/(2 c)$, we have 
$\frac{2}{3}\le \sqrt{a} A_n\le 2$.  Therefore, in this case the normalization constant remains bounded away from both zero and infinity, 
uniformly in $n$ and $\alpha$.  As a consequence of these estimates, we also have $\norm{\varphi_{n,0}-a^{-1/2}\phi_n}\le 2 c n/\alpha$.
}

To define $f(x,t)$ we first introduce the abbreviation
\begin{align}
 \beta = -\Im\, \tilde k \approx  \frac{n^2}{4a\alpha^2}
\end{align}
and recall
\be
 v = \frac{\hbar}{m} \Re\, \tilde k \approx \frac{\hbar\pi\alpha}{ma}\,,
\ee
whence $\tilde k = \frac{m}{\hbar} v-\ci \beta$ with $v,\beta>0$. We define further
\begin{align}
R(t)=a + v t, \qquad b(t) = \width^2 + \ci \frac{\hbar}{2 m} t ,
\end{align}
\xj{where the initial Gaussian spread $\width>0$ is left arbitrary for the moment (a convenient choice will turn out to be $\width= a$).}
The Gaussians will be attached symmetrically to 
$x=\pm R(t)$ with a ``variance'' $b(t)$, 
which yields explicitly
\begin{align}
f(x,t) = A_n \rme^{- \ci t Z_n/\hbar} \times \begin{cases} 
\pm B \rme^{\ci \tilde{k} (-x-a)-\frac{1}{4 b(t)} (-x-R(t))^2}, 
& \text{when } x<-R(t), \\
B \rme^{\ci \tilde{k} (x-a)-\frac{1}{4 b(t)} (x-R(t))^2}, 
&  \text{when } x>R(t),\\
\psi_n(x), & \text{when } |x|\le R(t).
 \end{cases}
\end{align}

Note that, for all $t\geq 0$, $f(\cdot,t)$ is normalizable but not normalized, and that $f(x,t)$ is continuously differentiable in $x$
because $\psi_n$ is, and because
the unnormalized Gaussian $\exp(-(x-\mu)^2/4 b)$ has, at its mean $\mu$,
value 1 and derivative 0. It is a short computation\footnote{%
The computation can be given the following mathematical justification:
Since the potential $V$ is bounded, by an application of the Kato--Rellich 
theorem \cite[Theorem X.15]{RS75}, 
the Hamiltonian  $H=-\tfrac{\hbar^2}{2m} \partial_x^2 + V$
is self-adjoint on the domain of $-\partial_x^2$.  It can be easily checked
that for any $t$ the derivative $\partial_x f(x,t)$ is absolutely continuous 
in $x$, and thus the function $f(\cdot,t)$ belongs to the domain of $H$.
This can be used to justify all the manipulations made here.  
Let us also use the 
opportunity to stress that, if we had not chosen the constants $A_\pm$ 
and $B_\pm$ in (\ref{eq:platansatz}) so that the function is continuously 
differentiable, then the addition of the Gaussian cut-off would have 
resulted in functions which are in $L^2(\RRR)$ but which {\em do not\/} 
belong to the domain of $H$.  Thus our 
estimates are {\em not\/} valid for such initial states.
For more sophisticated mathematical methods to study such problems,
see for instance \cite{skibsted86,CH09}.}
to check that for all $t> 0$
\begin{align}
 & (H-Z)f(x,t) = 
 -\tfrac{\hbar^2}{2m} \partial_x^2 f(x,t)+(V(x)-Z)f(x,t) \nonumber \\
 & =  -\tfrac{\hbar^2}{2m}
 \left[g_1(x-R(t),t) \pm g_1(-x-R(t),t) \right] f(x,t),
\end{align}
with (using the notation
$\chi(\cdots)$ as in \eqref{chidef})
\begin{align}
 g_1(y,t) = \chi(y>0) \left( \frac{y^2}{4 b(t)^2} 
   - \frac{1}{2 b(t)} - \ci \tilde{k} \frac{y}{b(t)}\right) .
\end{align}
In addition, we have
\begin{align}
 & \ci \hbar \partial_t f(x,t) = Z f(x,t) -
\tfrac{\hbar^2}{2m} \left[g_2(x-R(t),t) \pm g_2(-x-R(t),t) \right] f(x,t),
\end{align}
with $g_2=g_1+g_3$ where
\begin{align}
 g_3(y,t) = \chi(y>0) \frac{1+2 \beta y}{2 b(t)} .
\end{align}

As, for a fixed $t$, the function $f(x,t)$ is square integrable, we can
define a mapping
\begin{align}
 t\mapsto  F(t) = \rme^{\ci t H/\hbar} f(\cdot,t) - \varphi_0.
\end{align}
with $F(t)\in L^2$ for all $t\ge 0$, and
\be
\|F(0)\|^2 = \|f(\cdot,0)-\varphi_0\|^2
=\int_{a}^\infty \bigl|f(x,0) \bigr|^2 \rmd x +
\int_{-\infty}^{-a} \bigl| f(x,0) \bigr|^2 \rmd x\, .
\ee
\xj{
For any $t\ge 0$ and $|x|>R(t)$, the definition of $f$ yields
\begin{align}
|f(x,t)|^2 = |A_n|^2 |B|^2 \exp\!\left (2\, \Re\!\left[-\ci \frac{t}{\hbar} Z
  + \ci \tilde k (y + v t)- \frac{1}{4 b(t)} y^2\right] \right)\, ,
\end{align}
with $y=|x|-R(t)$.
Here the argument of the exponential can be simplified using 
$Z=\frac{\hbar^2}{2 m}\tilde k^2 -\Delta E$ to 
\begin{align}
 2\beta y-\frac{\width^2}{2 |b_t|^2} y^2 =
  \frac{1}{2} c_t^2 -   \frac{1}{2}
  \left[\frac{2\beta }{c_t}y - c_t\right]^2, \quad
  \text{with}\ c_t = 2 \beta |b(t)| \width^{-1}.
\end{align}
Thus for $t=0$, we have $c_0=2 \beta \width$ and $R(0)=a$, and by changing the integration variable to 
$y'=(|x|-a) 2\beta/c_0$, we find a bound
\begin{align}
\|F(0)\|^2 \le 2 |A_n|^2 |B|^2 \rme^{\frac{1}{2}c_0^2} \frac{c_0}{2 \beta} \int_0^\infty\!\rmd y'\, \rme^{-(y'-c_0)^2/2}
\le 2 \width |A_n|^2 |B|^2 \rme^{\frac{1}{2}c_0^2}\sqrt{2 \pi} \, .
\end{align}
Here $c_0\approx \frac{\sigma}{a} n^2/(2\alpha^{2})$, $A_n^2\approx 1/a$, and $|B|^2=|\kappa_n|^2=O(n^2\alpha^{-2})$.  
Thus if we choose $\sigma=a$, there is a pure constant $c'$ such that $\norm{F(0)}\le c' n/\alpha$ for all sufficiently small $n/\alpha$.
}

$F$ is differentiable and by the above estimates for all $t>0$, 
\begin{align}
 \partial_t F(t) = \rme^{\ci t H/\hbar} \left[ 
   \frac{\ci}{\hbar} H f(\cdot,t) + \partial_t f(\cdot,t) \right]
 = \rme^{\ci t H/\hbar} g(\cdot,t)
\end{align}
where
\begin{align}
 g(x,t) =  \ci \tfrac{\hbar}{2m} 
\left[g_3(x-R(t),t) \pm g_3(-x-R(t),t) \right] f(x,t) .
\end{align}
As the derivative is continuous (in the $L^2$-norm) in $t$, it can be
integrated to yield
$F(t) = F(0)+\int_0^t \rmd s\, \partial_s F(s)$.
Then, by the unitarity of the time evolution, \xj{we find
\begin{align}
\norm{f(\cdot,t)-\varphi_{t}} = \norm{F(t)} \le
\norm{F(0)} + \int_0^t \rmd s\, \norm{\partial_s F(s)} \le c' n \alpha^{-1}+
\int_0^t \rmd s\, \norm{g(\cdot,s)} .
\end{align}
}

\xj{
Thus we only need to estimate the magnitude of 
$\int_0^t \rmd s\, \norm{g(\cdot,s)}$.
As above,
}
\begin{align}
& \norm{g(\cdot,t)}^2  = \left(\frac{\hbar}{2 m}\right)^2 
2 \int_0^\infty \rmd y\, |f(y+R(t),t)|^2 |g_3(y,t)|^2
\nonumber \\ & \quad =
\left(\frac{\hbar |A_n| |B|}{2 m |b_t|}\right)^2
 \frac{1}{2} \int_0^\infty \rmd y\, (1 + 2 \beta y)^2 
 \exp\Bigl( 2\beta y-\frac{\width^2}{2 |b_t|^2} y^2\Bigr) 
\nonumber \\ & \quad =
\left(\frac{\hbar \beta |A_n| |B|}{m \width c_t}\right)^2
 \frac{c_t}{4 \beta} \rme^{\frac{1}{2} c_t^2}
\int_{-c_t}^\infty \rmd x\, (1 + c_t^2+ c_t x)^2 
\rme^{-\frac{1}{2} x^2}
 \nonumber\\ & \quad \le 
\left(\frac{\hbar \beta |A_n| |B|}{m \width c_t}\right)^2
 \frac{c_t}{4 \beta} \rme^{\frac{1}{2} c_t^2}
\int_{-\infty}^\infty \rmd x\, ((1 + c_t^2)^2 +  c_t^2 x^2) 
\rme^{-\frac{1}{2} x^2}
 \nonumber\\ & \quad =
\left(\frac{\hbar \sqrt{\beta} |A_n| |B|}{2 m \width}\right)^2
 \frac{1}{c_t} \rme^{\frac{1}{2} c_t^2} \sqrt{2 \pi}
 ((1 + c_t^2)^2 +  c_t^2) .
\end{align}
For sufficiently large $\alpha$ and all $0\le t\le\tau\approx (2ma^2/\hbar\pi n^2)\alpha$,
\begin{align}
c_t \leq c_{\tau} &= 2 \beta |b(\tau)| \width^{-1} = \frac{2}{\width}\beta\sqrt{\width^4 + \Bigl(\frac{\hbar}{2m}\Bigr)^2 \tau^2}\nonumber\\
&\approx \frac{2}{\width} \frac{n^2}{4a\alpha^2} \sqrt{\width^4 + \Bigl( \frac{a^2}{\pi n^2}\Bigr)^2\alpha^2}\leq \frac{a}{\pi\width} \frac{1}{\alpha}\,,
\end{align}
\xj{and therefore then}
\begin{align}
& \norm{g(\cdot,t)}  \le
\frac{\hbar \sqrt{\beta} |A_n| |B|}{2 m \width}
 \frac{2}{\sqrt{c_t}}\,.
\end{align}
Since
\begin{align}
  c_s = \sqrt{(2 \beta \width)^2 + (s \beta \hbar/(m\width))^2} \ge 
s \frac{\beta \hbar}{m\width},
\end{align}
we can estimate the integral over $s$ by
\begin{align}
\int_0^t \rmd s\, \frac{1}{\sqrt{c_s}} \le
\int_0^t \rmd s\, \sqrt{\frac{m \width}{\beta \hbar s}} 
= 2 \sqrt{\frac{m \width t}{\beta \hbar}}  .
\end{align}
\xj{This proves that for all $0\le t\le \tau$, and sufficiently small $n/\alpha$}
\begin{align}\label{eq:Ffinalest}
& \norm{f(\cdot,t)-\varphi_{t}}^2 \le
2\|F(0)\|^2 +8|A_n|^2 |B|^2 \frac{\hbar}{m \width} t 
\le 2 (c')^2 \frac{n^2}{\alpha^2}
+\frac{4a}{\pi\width}  \frac{t}{\tau}\frac{1}{\alpha} \ll 1\,,
\end{align} 
where we have used $(|z|+|z'|)^2\le 2 (|z|^2+|z'|^2)$, valid for all $z,z'\in \CCC$ by H\"older's inequality.
Since on the interval $[-a-vt,a+vt]$, $f(x,t) = A_n \psi_{n,t}(x)$, we have that
\be
\int_{-a-vt}^{a+vt} |\varphi_{t}(x) - A_n\psi_{n,t}(x)|^2 \rmd x \leq \int_{-\infty}^\infty
|\varphi_{t}(x) - f(x,t)|^2 \rmd x = \norm{f(\cdot,t)-\varphi_{t}}^2 \ll 1\,,
\ee
which is what we wanted to show.\hfill$\square$

\bigskip

\bigskip

\noindent\textit{Proof of Corollary \ref{corr:metastable}.} It follows easily from Theorem~\ref{thm:metastable}: Writing $\varphi_{n,0}$ for the wave function in \eqref{varphi0}, we have that
\be\label{psivarphi}
\psi(x) = \sum_{n=1}^{n_{\mathrm{max}}} \frac{c_n}{A_n} \varphi_{n,0}(x)\,.
\ee
From \eqref{psivarphi} we obtain that, provided $0<t<\tau_n$ for each $n$,
\begin{align}
&\Bigl\|\Bigl(\psi_t- \sum_n c_n \psi_{n,t}\Bigr) \chi(-a-vt\le x \le a+vt) \Bigr\|\nonumber \\
\le& \sum_n \Bigl|\frac{c_n}{A_n}\Bigr| \, \Bigl\|\Bigl( \varphi_{n,t}- A_n \psi_{n,t} \Bigr) \chi(-a-vt\le x \le a+vt)  \Bigr\| \ll 1
\end{align}
with the notation $\chi(\cdots)$ as in \eqref{chidef}.  
This proves \eqref{combcomp}.\hfill$\square$

\bigskip\bigskip

\noindent\textit{Proof of Theorem~\ref{thm:confinement1}.}
As we proved above, for all small enough 
$n/\alpha$ the vectors $\varphi_{n,0}$ can be approximated by $e_n(x)=\pm
a^{-1/2}\sin(\frac{\pi n}{2 a} (x+a))$ with the error bounded by $c n/\alpha$, 
$c$ a numerical constant.
The functions $e_n$ are up to a sign equal to the sine-basis of square
integrable functions on $[-a,a]$, and therefore they form an orthonormal
basis. Let $a_n$ denote the expansion constants of $\psi_0$ in this basis, i.e., they are the unique constants for which 
$\psi_0=\sum_{n=1}^\infty a_n e_n$.  Since $a_n$
are obtained by projecting $\psi_0$ to $e_n$, they depend only on $\psi_0$, $a$, and $n$.

Now $\sum_n |a_n|^2 = \norm{\psi_0}^2 = 1$, and for any given $\vep$,
there is an $\alpha$-independent constant $n_\mathrm{max}(\vep)<\infty$, such that 
\begin{align}
 \Bigl\Vert\psi_0-\sum_{n=1}^{n_\mathrm{max} (\vep) } a_n
e_{n}\Bigr\Vert  \le \frac{1}{4} \vep\, .
\end{align}
Also, necessarily
$\sum_{n=1}^{n_\mathrm{max} (\vep) } |a_n|^2 \ge 1-\vep^2/16$. Therefore,
\begin{align}
 \Bigl\Vert\psi_0-\sum_{n=1}^{n_\mathrm{max} (\vep) }a_n
\varphi_{n,0}\Bigr\Vert  \le \frac{1}{4} \vep + \frac{c
n_\mathrm{max}(\vep)^2}{\alpha}\, .
\end{align}

As we proved in Appendix \ref{sec:plateigen}, for any fixed $n$, $\tau_n\to
\infty$ in the limit $\alpha\to\infty$.  Therefore, for all sufficiently
large $\alpha$, we \y{have} $t_0\le \tau_n$, for all $1\le n\le n_\mathrm{max}
(\vep)$.
Thus by the explicit estimate in \eqref{eq:Ffinalest}, the time-evolved
\y{vectors in Hilbert space}
satisfy for such large $\alpha$ \y{and any $0\le t \le t_0$}
\begin{align}
  & \Bigl\Vert\chi(|x| \le a) \, \Bigl(
 \psi_t-\sum_{n=1}^{n_\mathrm{max} (\vep)}
a_n A_n \psi_{n,t} \Bigr)\Bigr\Vert 
\\ \nonumber & \quad 
\le \Bigl\Vert
\psi_t-\sum_{n=1}^{n_\mathrm{max} (\vep) }a_n \varphi_{n,t}\Bigr\Vert 
+ \sum_{n=1}^{n_\mathrm{max}(\vep)} |a_n| \Bigl\Vert \chi(|x| \le a) \, \Bigl(
\varphi_{n,t}-A_n\psi_{n,t}\Bigr)\Bigr\Vert
\\ \nonumber & \quad 
\le \frac{1}{4} \vep + \frac{c
n_\mathrm{max}(\vep)^2}{\alpha} + c'\frac{n_\mathrm{max}(\vep)^3}{\alpha^2} 
\, ,
\end{align}
with some numerical constant $c'$.  For sufficiently large $\alpha$, the
right hand side is \y{bounded by $\vep/2$.}

Since $A_n \psi_{n,t}(x)\chi(|x| \le a) = 
\rme^{-\ci Z_n t /\hbar}\varphi_{n,0}(x)$ and
$\lim_{\alpha\to \infty} Z_n = \frac{\hbar^2 \pi^2 n^2}{8 m a^2}$, 
we have $A_n \psi_{n,t}(x)\chi(|x| \le a) \to 
\rme^{-\ci\frac{\hbar \pi^2 n^2}{8 m a^2} t}e_{n}(x)$ in norm 
when $\alpha\to\infty$, \y{in fact uniformly in $t\in [0,t_0]$.}  This implies that
\begin{align}
 \lim_{\alpha\to\infty}\Bigl\Vert\chi(|x| \le a)  \sum_{n=1}^{n_\mathrm{max}
(\vep) }a_n A_n \psi_{n,t}\Bigr\Vert^2
= \sum_{n=1}^{n_\mathrm{max} (\vep) }|a_n|^2 \ge 1- \frac{\vep^2}{16}
\end{align}
\y{uniformly in $t\in[0,t_0]$, and thus}
\be
\Bigl\Vert\chi(|x| \le a)  \sum_{n=1}^{n_\mathrm{max}
(\vep) }a_n A_n \psi_{n,t}\Bigr\Vert^2 \ge 1- \frac{\vep^2}{8}
\ee 
\y{for all $t\in [0,t_0]$, provided $\alpha$ is big enough. By the triangle inequality, 
$\norm{\chi(|x| \le a) \psi_t}\ge \sqrt{1-\vep^2/8}-\vep/2$.} As $\norm{\psi_t}=1$, then necessarily 
$\norm{\chi(|x| > a) \psi_t}^2\le 1-(\sqrt{1-\vep^2/8}-\vep/2)^2< \vep$.
\y{This} concludes the proof of Theorem~\ref{thm:confinement1}.\hfill$\square$

\bigskip\bigskip

\noindent \textit{Acknowledgments.} We thank the Institut des Hautes
\'Etudes Scientifiques at Bures-sur-Yvette, France, where the idea for this
article was conceived, for hospitality. For discussions on the topic we
thank in particular Federico Bonetto (Georgia Tech, USA), Ovidiu Costin
(Ohio State University), and Herbert Spohn (TU M\"unchen, Germany). 

The work of S.~Goldstein was supported in part by NSF Grant DMS-0504504.
The work of J.~Lukkarinen was supported by the Academy of Finland and by the 
Deutsche Forschungsgemeinschaft (DFG) project Sp~181/19-2. 
The work of R.~Tumulka was supported by the European Commission
through its 6th Framework Programme ``Structuring the European
Research Area'' and the contract Nr. RITA-CT-2004-505493 for the
provision of Transnational Access implemented as Specific Support
Action.

\end{document}